\documentclass[twocolumn]{article}
\usepackage{cite}
\PassOptionsToPackage{hyphens}{url}
\usepackage{hyperref}
\usepackage{graphicx}
\usepackage{tabularx}
\usepackage{booktabs}
\usepackage{authblk}
\usepackage{abstract}
\usepackage{mathbbol}
\usepackage{longtable}
\usepackage{geometry}

\title{\textbf{Mapping the Intellectual Structure of Social Network Research: A Comparative Bibliometric Analysis}}
\author[1]{Pengjia Cui \thanks{Corresponding author: pcui@ucsd.edu}}
\author[2]{Yawen Dong}
\affil[1,2]{Computational Social Science, 
	University of California San Diego}
\date{}



\begin{document}
	
	\twocolumn[
	\maketitle
	\begin{onecolabstract}
		Network science is an interdisciplinary field that transcends traditional academic boundaries, offering profound insights into complex systems across disciplines. This study conducts a bibliometric analysis of three leading journals—\textit{Social Networks}, \textit{Network Science}, and the \textit{Journal of Complex Networks}—each representing a distinct yet interconnected perspective within the field. \textit{Social Networks} focuses on empirical and theoretical advancements in social structures, emphasizing sociological and behavioral approaches. \textit{Network Science} bridges physics, computer science, and applied mathematics to explore network dynamics in diverse domains. The \textit{Journal of Complex Networks}, by contrast, is dedicated to the mathematical and algorithmic foundations of network theory. By employing co-authorship and citation network analysis, we map the intellectual landscape of these journals, identifying key contributors, influential works, and structural trends in collaboration. Through centrality measures such as degree, betweenness, and eigenvector centrality, we uncover the most impactful publications and their roles in shaping the discourse within and beyond their respective domains. Our analysis not only delineates the disciplinary contours of network science but also highlights its convergence points, revealing the evolving trajectory of this dynamic and rapidly expanding field.
		
	\end{onecolabstract}
	
	
	\vspace{1cm}
	]

	\section{Introduction}\label{Introduction}
	
	Bibliometric analysis provides a systematic method for uncovering the intellectual structure and research dynamics within a scientific field \cite{donthu2021bibliometric,arias2020bibliometric, WOS:001124720500001, WOS:000208374100011}. It enables the identification of key research themes, methodological trends, and the evolution of scholarly discourse. As an inherently interdisciplinary domain, network science—spanning sociology, computer science, mathematics, and physics—poses unique challenges for bibliometric studies due to its methodological and application diversity \cite{Newman2018,Barabasi2016}.
	
	This study analyzes three leading journals in the field: \textit{Social Networks}, \textit{Journal of Complex Networks}, and \textit{Network Science}. These journals were selected for their distinct scopes: \textit{Social Networks} emphasizes sociological and empirical studies \cite{wasserman1994social}, \textit{Journal of Complex Networks} focuses on mathematical and computational approaches \cite{boccaletti2006complex}, and \textit{Network Science} bridges interdisciplinary perspectives \cite{Newman2010}. By examining these journals, we aim to provide a comprehensive overview of the intellectual structure, thematic evolution, and collaboration patterns within social and complex network research.
	
	Our objectives are to: (1) map the intellectual structure of the field via citation networks and scholarly influences \cite{Small1973}, (2) identify key themes and trends through keyword co-occurrence and thematic clustering \cite{callon1983co}, (3) evaluate research impact by analyzing citations, influential publications, and prolific contributors \cite{garfield1979citation}, and (4) explore collaboration patterns through co-authorship and institutional networks \cite{glanzel2004coauthorship}.
	
	To achieve these goals, we employ a three-pronged approach. First, we conduct a performance analysis to assess publication trends, citation impacts, and contributions of leading scholars and institutions \cite{bornmann2011citation}. Second, we utilize science mapping techniques, including co-citation analysis, bibliographic coupling, and keyword co-occurrence, to uncover thematic clusters and intellectual structures \cite{van2010visualizing,chen2017citespace}. Third, we perform network analysis to examine collaboration patterns across authors, institutions, and countries \cite{newman2004coauthorship, moody2004sociology}.
	
	By integrating these methods, this study offers a structured, data-driven overview of the research landscape in social and complex network studies. It highlights the field's development, emerging trajectories, and potential future directions.

	\section{Bibliometric Techniques}\label{Bibliometric Techniques}
	
	Table \ref{table.tab2} presents a comparative analysis of key metrics across the three journals, including publication frequency, impact factors, citation patterns, and collaboration trends\cite{donthu2021bibliometric}. The table highlights both shared and distinguishing features, such as publisher affiliations, research focus, and the thematic emphasis of highly cited works. This comparison provides a structured overview of each journal’s influence and role within social and complex network research\cite{ WOS:000356343600002}.
	
	\begin{table*}[htbp]
		\tiny
		\caption{Comparison of Metrics}\label{table.tab2}
		\begin{tabularx}{\textwidth}{|p{5cm}|X|X|X|}
			\hline
			\textbf{Category} & \textbf{Social Networks} & \textbf{Network Science} & \textbf{Journal of Complex Networks} \\ \hline
			\textbf{Publisher} & Elsevier & Cambridge University Press & Oxford University Press \\ \hline
			\textbf{Publication Frequency} & 4 issues/year & 4 issues/year & 6 issues/year \\ \hline
			\textbf{Impact Factor (2023)} & 2.9 & 1.4 & 2.2 \\ \hline
			\textbf{5-Year Impact Factor} & 3.1 & 1.7 & 2.0 \\ \hline
			\textbf{Eigenfactor Score} & 0.00387 & 0.00084 & 0.00160 \\ \hline
			\textbf{Article Influence Score} & 1.368 & 0.741 & 0.692 \\ \hline
			\textbf{Cited Half-life (years)} & 16.3 & 7.4 & 5.7 \\ \hline
			\textbf{JCI (2023)} & 1.89 & 0.66 & 0.63 \\ \hline
			\textbf{Immediacy Index} & 0.8 & 0.2 & 0.5 \\ \hline
			\textbf{Total Citable Items (2023)} & 240 & 99 & 163 \\ \hline
			\textbf{Open Access Percentage} & 31.67\% & 44.44\% & 22.09\% \\ \hline
			\textbf{Top Contributing Organization} & University of Oxford (14) & University of California (12) & Santa Fe Institute (7) \\ \hline
			\textbf{Top Contributing Country} & USA (103 papers) & USA (59 papers) & USA (39 papers) \\ \hline
			\textbf{Total Citations (2023)} & 6,795 & 856 & 1,117 \\ \hline
			\textbf{Self-citations (\%)} & 5.55\% & 3.62\% & 3.58\% \\ \hline
			\textbf{Top Citation Source} & Social Networks (43) & Social Networks (34) & Physica A (35) \\ \hline
			\textbf{Top Cited Source} & Social Networks (377) & PNAS (51) & Physical Review E (159) \\ \hline
			\textbf{International Collaboration (\%)} & 47.2\% & 51.3\% & 45.6\% \\ \hline
			\textbf{Top Collaborating Countries} & USA-England & USA-Germany & USA-China \\ \hline
			\textbf{High-frequency Keywords} & Social capital, community & Network topology, algorithms & COVID-19, resilience \\ \hline
			\textbf{Emerging Topics} & Computational sociology & Complex systems & Epidemic modeling \\ \hline
			\textbf{Annual Growth Rate (2010-2023)} & +6.5\% & +8.2\% & +5.4\% \\ \hline
		\end{tabularx}	
		
	\end{table*}
	
	To systematically analyze the intellectual structure of social network research, this study employs three interconnected bibliometric approaches: \textbf{performance analysis}, \textbf{science mapping}, and \textbf{network analysis}.\cite{donthu2021bibliometric} Each method provides distinct yet complementary insights into the research landscape of the selected journals.
	
	\textbf{Performance analysis} quantitatively evaluates research productivity and impact by examining publication trends, citation metrics, and contributions from influential authors, institutions, and countries\cite{ WOS:000331771900008}. This approach highlights the historical growth of the field and identifies its most prolific contributors.
	
	\textbf{Science mapping} explores the thematic and conceptual structure of the field by uncovering research clusters, keyword co-occurrence patterns, and intellectual influences \cite{Small1973, Callon1991, Borner2003}. Using techniques such as co-citation analysis \cite{Small1973}, bibliographic coupling \cite{Kessler1963}, and topic modeling \cite{Blei2003}, science mapping reveals the knowledge base and thematic evolution of social network research \cite{Chen2006, Cobo2011}.
	
	\textbf{Network analysis} investigates collaboration patterns among authors, institutions, and countries. By analyzing co-authorship networks, institutional collaborations, and international partnerships, this approach sheds light on the social and structural dynamics of knowledge production in the field\cite{ WOS:000336729500129}.
	
	By systematically integrating these three analytical components, this study provides a comprehensive and structured overview of social network research\cite{donthu2021bibliometric,yazdanjue_comprehensive_2023}. The results will:
	\begin{itemize}
		\item Reveal historical and emerging research trends, highlighting shifts in dominant themes and methodological approaches.
		\item Identify key contributors and influential works, offering insights into the most impactful papers, prolific scholars, and leading institutions.
		\item Map collaborative networks, examining how scholars, institutions, and countries interact and contribute to the field.
	\end{itemize}
	This multi-faceted approach not only quantifies research impact but also contextualizes the development of ideas, theories, and methodologies in social network research \cite{Small1973, Callon1991,Cobo2011}. By integrating performance analysis, science mapping, and network analysis, we offer a comprehensive examination of the field’s intellectual structure and collaborative dynamics. The following sections detail each bibliometric technique, outlining its objectives, methodological framework, data requirements, and analytical tools.
	
	To ensure the robustness and accuracy of our bibliometric dataset, we retrieved publication records from the Web of Science (WoS), focusing exclusively on articles published in \textit{Social Networks}, \textit{Journal of Complex Networks}, and \textit{Network Science}. The search was carefully structured to include only publications explicitly indexed under these journals, thereby minimizing misclassification and ensuring dataset integrity \cite{Moed2005}. The full query links used for data extraction are provided below:
	\begin{itemize}
		\item \url{https://www.webofscience.com/wos/woscc/summary/8b9cb9e7-5920-4a8c-8957-1f45746eb38f-01449729f7/relevance/1}
		\item 	\url{https://www.webofscience.com/wos/woscc/summary/b70d5df8-5cd9-4064-8329-390221e5fcc0-014497387b/relevance/1}
		\item 	\url{https://www.webofscience.com/wos/woscc/summary/cfb2985d-4457-4214-85c1-4ee6224ffecb-014497425b/relevance/1}
	\end{itemize}
	
	Following data retrieval, a preprocessing step was conducted to refine the dataset and ensure data integrity. First, we verified that each record was explicitly affiliated with one of the three target journals by cross-referencing journal names with standardized indexing metadata.\cite{donthu2021bibliometric}. Entries with inconsistencies—such as incorrect journal attributions, duplicate records, or misclassified publication types—were removed. Additionally, records with incomplete metadata, including missing publication years, author information, or improperly formatted citation fields, were excluded \cite{Aria2017}. To further enhance accuracy, we employed VOSviewer for citation data validation and cleaning, ensuring a robust dataset for analysis \cite{vanEck2010}. These measures minimized distortions and ensured that our bibliometric analysis accurately captured the scholarly output of each journal.

	\section{Performance Analysis}\label{Performance Analysis}
	
	Performance analysis in bibliometrics provides an overview of research productivity, citation impact, and contributions of authors, institutions, and countries. It helps quantify the most influential publications, prolific researchers, and overall research trends in \textit{Social Networks}, \textit{Journal of Complex Networks}, and \textit{Network Science}.
	
	\subsection{Publication Trends in Social Network Research}
	
	\begin{figure}[htbp]
		\centering
		\caption{Publication Trends}\label{fig.fig1}
		\includegraphics[width=\columnwidth]{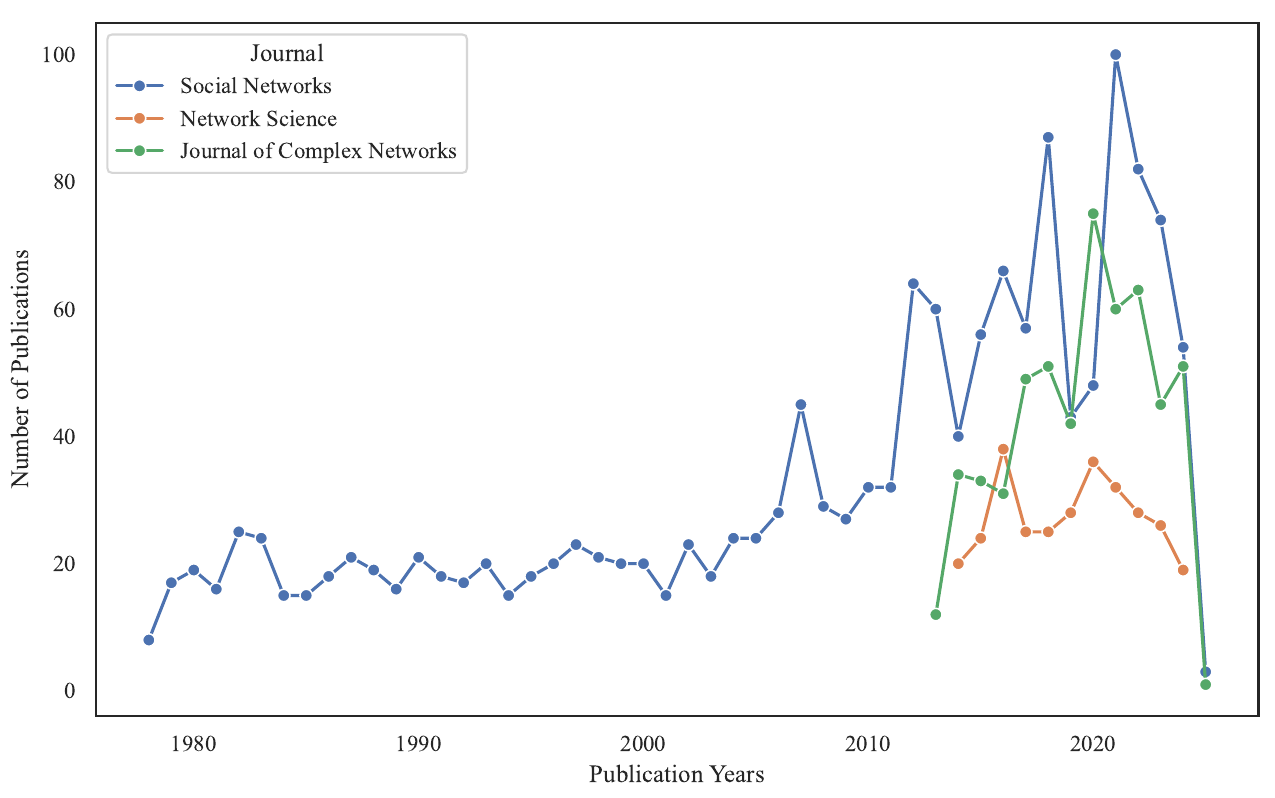}
	\end{figure}
	
	Examining the publication trajectories of \textit{Social Networks}, \textit{Journal of Complex Networks}, and \textit{Network Science} provides insight into the evolution of social network research. Figure \ref{fig.fig1} illustrates how research output has changed over time, highlighting the growth and specialization of the field.
	
	\subsubsection*{Expansion and Institutionalization of Social Network Research}
	
	\textit{Social Networks}, the longest-running journal in the field, has published research continuously since its founding in 1978 \cite{Freeman2004}. For decades, it served as the primary outlet for social network analysis, maintaining stable publication volumes. However, since 2010, a notable increase in output reflects the growing influence of computational methods and empirical applications \cite{Borgatti2009}. This trend suggests a broadening of the field as new methodologies and interdisciplinary collaborations expand its research scope.
	
	The establishment of \textit{Journal of Complex Networks} in 2013 marked the rising prominence of mathematical and computational approaches to network analysis \cite{Barabasi2016}. Initially publishing at a modest rate, its output grew rapidly after 2015, nearing that of \textit{Social Networks} by 2020. This growth highlights the emergence of complex network research as a distinct subfield, attracting contributions from applied mathematics, computer science, and physics \cite{Borgatti2009}.
	
	\textit{Network Science}, introduced in 2014, maintains a comparatively lower publication volume, indicating its focus on a specialized academic community. Emphasizing foundational principles and interdisciplinary perspectives, the journal has experienced steady growth but remains the smallest of the three in annual output, reinforcing its niche within the broader field of network science \cite{Barabasi2016}.

	\subsubsection*{Publication Growth and Recent Trends}
	
	Between 2016 and 2022, publication activity in all three journals increased significantly, reflecting the growing influence of network science \cite{Newman2018, Barabasi2016}. \textit{Social Networks} exceeded 100 publications per year, while \textit{Journal of Complex Networks} and \textit{Network Science} followed similar trajectories on a smaller scale \cite{Borgatti2013}. This growth aligns with the rise of big data, computational social science, and machine learning, which have reinforced the prominence of network-based methodologies \cite{Lazer2009, Watts2011}.
	
	Since 2022, publication volumes have stabilized or declined, potentially signaling research saturation or database indexing delays \cite{Fortunato2018}. Further analysis is needed to determine whether this trend reflects a plateau or a shift in thematic focus.

	\subsubsection*{Thematic Differentiation Among Journals}
	
	The trajectories of these journals highlight the thematic specialization within social network research. \textit{Social Networks} remains the primary outlet for applied and empirical studies, emphasizing sociology, organizational research, and human behavior. \textit{Journal of Complex Networks} focuses on computational and mathematical approaches, while \textit{Network Science} bridges theoretical and interdisciplinary perspectives.
	
	This specialization reflects the diversification of network science into distinct subfields, each catering to specific research communities. Further analysis of citation networks and thematic clustering will clarify how these journals interact and shape the field over time.
	
	\subsection{Prolific Authors}\label{Prolific Authors}
	
	This section investigates the contributions of the most prolific authors within the domain of social network research, across three prominent journals: \textit{Social Networks}, \textit{Journal of Complex Networks}, and \textit{Network Science}. By analyzing the publication records, we uncover the significant roles these individuals play in shaping the field and defining its intellectual boundaries.
	
	\begin{figure*}[htbp]
		\centering
		\includegraphics[width=\textwidth]{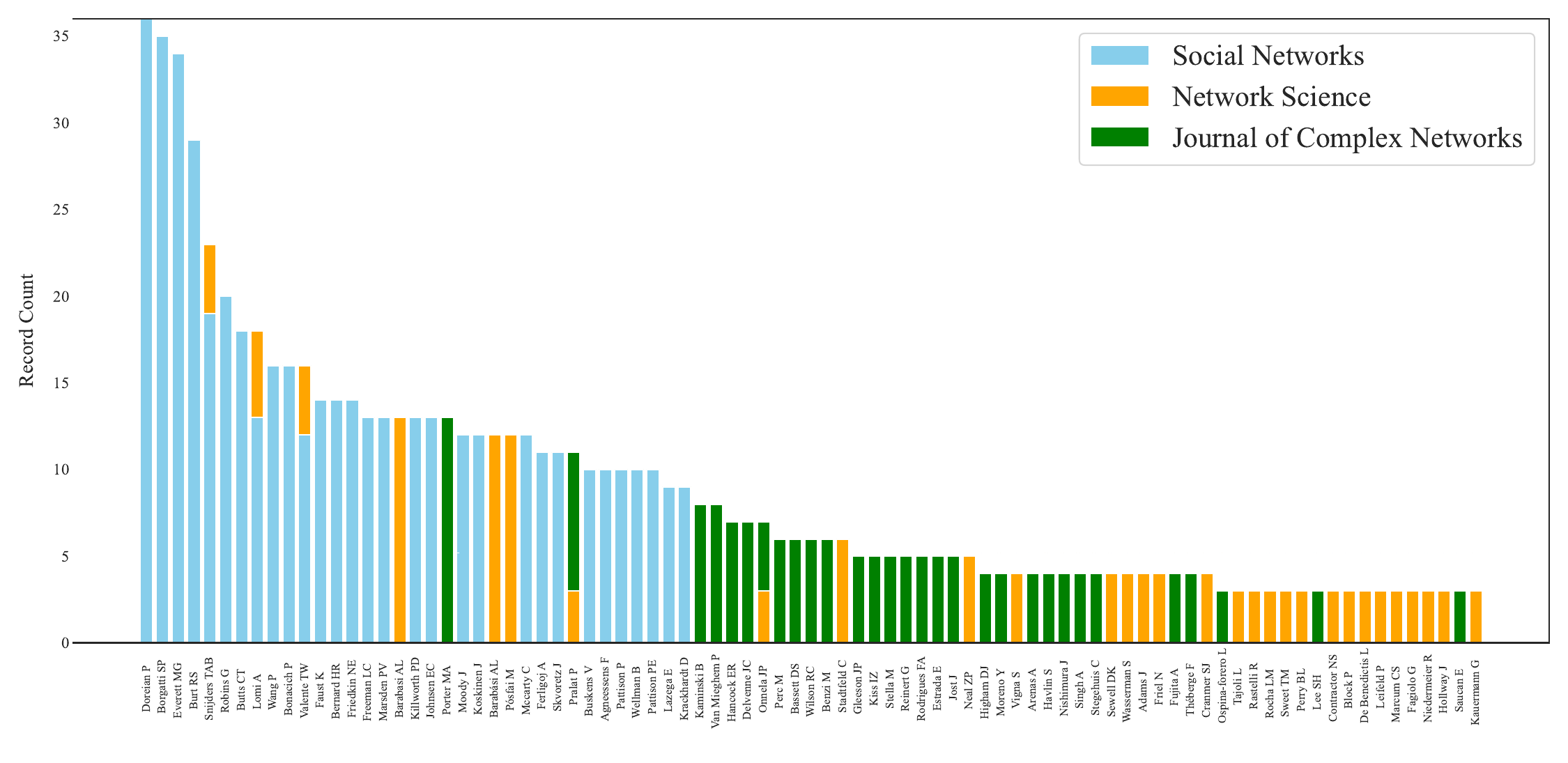}
		\caption{Prolific Authors}
		\label{fig.fig2}
	\end{figure*}
	
	\subsubsection*{Author Contributions and Research Orientations}
	
The trajectories of these journals highlight the thematic specialization within social network research. \textit{Social Networks} remains the primary outlet for applied and empirical studies, emphasizing sociology, organizational research, and human behavior \cite{Marin2011,Freeman2004}. \textit{Journal of Complex Networks} focuses on computational and mathematical approaches, reflecting its strong association with statistical physics, graph theory, and algorithmic network analysis \cite{Newman2010,Fortunato2010}. Meanwhile, \textit{Network Science} bridges theoretical and interdisciplinary perspectives, integrating social, biological, and technological networks under a unified framework \cite{Barabasi2022,Boccaletti2006}.

This specialization reflects the diversification of network science into distinct subfields, each catering to specific research communities. \textit{Social Networks} is rooted in social sciences and policy-relevant research, whereas \textit{Journal of Complex Networks} prioritizes methodological advancements in network modeling. \textit{Network Science}, by contrast, functions as a hybrid platform promoting interdisciplinary synthesis \cite{Lazer2009,Borner2020}. Further analysis of citation networks and thematic clustering will clarify how these journals interact and shape the field over time.

	\subsubsection*{Patterns of Specialization and Cross-Journal Engagement}
	
	The limited cross-journal publication by authors may stem from the distinct academic cultures and publication strategies inherent in their areas of expertise. It is reasonable to hypothesize that:
	\begin{itemize}
		\item \textbf{Empirical and Applied Research Focus:} Authors publishing predominantly in \textit{Social Networks} might prioritize empirical data and real-world applications, which aligns with the journal's aim to influence practical and policy-related outcomes.
		\item \textbf{Theoretical and Computational Focus:} Conversely, authors like \textbf{Porter MA} and \textbf{Barabási AL} engage with journals like \textit{Journal of Complex Networks} and \textit{Network Science} due to their interest in developing new theoretical frameworks and computational models that may not align with the more applied nature of \textit{Social Networks}.
	\end{itemize}
	
	This specialization underlines a broader academic phenomenon where researchers often become siloed within their disciplinary boundaries, occasionally leading to challenges in interdisciplinary research dissemination \cite{Leahey2016,Porter2009}. The fact that very few authors publish across all three journals suggests a significant opportunity for promoting interdisciplinary research, which could bridge gaps between empirical and computational studies \cite{Wagner2011,Chavarro2017}. Encouraging cross-disciplinary collaboration may facilitate methodological integration and foster novel insights at the intersection of social, mathematical, and computational network analysis.

	\subsubsection*{Implications for the Field}
	
	This segmentation of publishing within specific journals reflects broader intellectual trends and may indicate potential barriers to interdisciplinary research \cite{Leahey2016,Porter2009}. While \textit{Social Networks} remains a primary venue for empirical research, theoretical advancements published in \textit{Journal of Complex Networks} and \textit{Network Science} could provide valuable frameworks for refining empirical models and expanding methodological approaches \cite{Chavarro2017,Wagner2011}.
	
	The presence of a small but notable group of cross-journal contributors suggests a pathway toward greater integration of computational and empirical approaches. Encouraging interdisciplinary collaboration could facilitate methodological synthesis, enabling computational models to complement empirical investigations and leading to a more comprehensive understanding of network structures and dynamics \cite{Rafols2010,VanNoorden2015}.
	
	In subsequent sections, we explore how these publication patterns shape the intellectual landscape of network research and assess their implications for the integration of methodological innovations across disciplines.

	\subsection{Institutional Contributions to Social Network Research}\label{Institutional Contributions to Social Network Research}
	
	This section examines the leading institutional contributors to social network research, focusing on their publication records across three key journals: \textit{Social Networks}, \textit{Journal of Complex Networks}, and \textit{Network Science}. By analyzing the top affiliations in each journal, we gain insight into dominant institutions, regional patterns, and the differing research orientations reflected in these publication venues \cite{Wagner2018, Glanzel2015, Leydesdorff2021}.
	
	\subsubsection*{Institutional Influence and Regional Disparities}
	
	A clear pattern emerges in the institutional distribution of social network research. The \textit{University of California System} is the most prolific contributor, with substantial publication output across all three journals. Its presence is most pronounced in \textit{Social Networks}, where it accounts for the largest institutional share, but it also maintains a notable footprint in \textit{Journal of Complex Networks} and \textit{Network Science}. This broad engagement underscores its commitment to both empirical and computational network science.
	
	Beyond the University of California System, North American institutions dominate contributions to \textit{Social Networks}, reinforcing the journal’s strong ties to sociology and applied network analysis \cite{moody2004sociology, Freeman2004}. Universities such as \textit{Pennsylvania Commonwealth System of Higher Education}, \textit{University of Pittsburgh}, and \textit{University of California Irvine} rank among the most frequent contributors. Meanwhile, European universities, particularly \textit{University of Groningen} and \textit{University of Oxford}, play a central role, highlighting the journal’s reach beyond the United States \cite{Borgatti2009}.
	
	\subsubsection*{Theoretical Focus in Journal of Complex Networks}
	
	In contrast, \textit{Journal of Complex Networks} features a stronger presence of European institutions, reflecting its emphasis on mathematical and algorithmic approaches to network science \cite{Newman2010}. The leading contributors include \textit{University of Oxford}, \textit{Centre National de la Recherche Scientifique (CNRS)}, and \textit{University of London}, institutions known for their focus on theoretical modeling and complexity science \cite{Barabasi2016}. Many of these affiliations maintain collaborations with physics and computer science departments, further reinforcing the journal’s orientation toward formal network analysis.
	
	\subsubsection*{Interdisciplinary Engagement in Network Science}
	
	\textit{Network Science} presents a more interdisciplinary institutional composition, incorporating both theoretical and applied perspectives. It attracts contributions from leading research universities, including \textit{Harvard University}, \textit{Indiana University}, and \textit{Central European University}, each of which has established itself as a center for computational and quantitative social science \cite{Fortunato2018}. Additionally, institutions such as \textit{The Santa Fe Institute} and \textit{CNRS} are well represented, reflecting the journal’s emphasis on interdisciplinary and fundamental research in network theory \cite{Boccaletti2014}.
	
	\subsubsection*{Institutional Overlap and Specialization}
	
	While a few institutions maintain a presence across all three journals, most exhibit specialization in either applied or theoretical network research. Universities such as \textit{Oxford} and \textit{California} contribute broadly, spanning both empirical and computational network studies \cite{Strogatz2001}. However, others, like \textit{CNRS} and \textit{Harvard}, are more concentrated in \textit{Journal of Complex Networks} and \textit{Network Science}, respectively, signaling a stronger focus on formal network methodologies. The relative lack of institutional overlap suggests that, despite their shared focus on network research, these journals cater to distinct scholarly communities.
	
	\subsubsection*{Implications for the Field}
	
	The institutional landscape of social network research reflects both regional and disciplinary distinctions. \textit{Social Networks} remains closely linked to North American universities with strong traditions in empirical network studies, while European institutions lead contributions to \textit{Journal of Complex Networks} and \textit{Network Science}, reinforcing their prominence in complexity science and theoretical research \cite{Watts1998}. The presence of highly specialized institutions such as \textit{The Santa Fe Institute} highlights the growing role of interdisciplinary approaches, while increasing contributions from regions outside North America and Europe—such as \textit{Universidade de São Paulo}—signal a gradual globalization of the field.

	\begin{table}[htbp]
		\centering
		\caption{Top 10 Affiliations per Journal}
		\label{table.tab1}
		\resizebox{\columnwidth}{!}{
			\begin{tabular}{lcl}
				\toprule
				Affiliations & Record Count & Journal \\
				\midrule
				UNIVERSITY OF CALIFORNIA SYSTEM               & 171 & Social Networks \\
				UNIVERSITY OF CALIFORNIA IRVINE               & 86 & Social Networks \\
				PENNSYLVANIA COMMONWEALTH SYSTEM OF HIGHER ED & 73 & Social Networks \\
				UNIVERSITY OF GRONINGEN                       & 66 & Social Networks \\
				UNIVERSITY OF OXFORD                          & 55 & Social Networks \\
				UNIVERSITY OF PITTSBURGH                      & 52 & Social Networks \\
				UTRECHT UNIVERSITY                            & 47 & Social Networks \\
				UNIVERSITY OF CHICAGO                         & 41 & Social Networks \\
				UNIVERSITY OF MELBOURNE                       & 41 & Social Networks \\
				UNIVERSITY OF SOUTH CAROLINA COLUMBIA         & 41 & Social Networks \\
				UNIVERSITY OF OXFORD                          & 30 & Journal of Complex Networks \\
				UNIVERSITY OF CALIFORNIA SYSTEM               & 21 & Journal of Complex Networks \\
				HARVARD UNIVERSITY                            & 20 & Network Science \\
				UNIVERSITY OF CALIFORNIA SYSTEM               & 18 & Network Science \\
				CENTRE NATIONAL DE LA RECHERCHE SCIENTIFIQUE  & 18 & Journal of Complex Networks \\
				UNIVERSITY OF LONDON                          & 16 & Journal of Complex Networks \\
				INDIANA UNIVERSITY BLOOMINGTON                & 15 & Network Science \\
				INDIANA UNIVERSITY SYSTEM                     & 15 & Network Science \\
				NORTHEASTERN UNIVERSITY                       & 15 & Network Science \\
				CENTRAL EUROPEAN UNIVERSITY                   & 14 & Network Science \\
				UNIVERSIDADE DE SAO PAULO                     & 14 & Journal of Complex Networks \\
				CENTRE NATIONAL DE LA RECHERCHE SCIENTIFIQUE  & 13 & Network Science \\
				THE SANTA FE INSTITUTE                        & 13 & Journal of Complex Networks \\
				HARVARD MEDICAL SCHOOL                        & 12 & Network Science \\
				UNIVERSITY OF OXFORD                          & 12 & Network Science \\
				MAX PLANCK SOCIETY                            & 11 & Journal of Complex Networks \\
				UNIVERSITY OF CALIFORNIA LOS ANGELES          & 11 & Journal of Complex Networks \\
				UNIVERSITY OF TRENTO                          & 11 & Journal of Complex Networks \\
				UNIVERSITY OF GRONINGEN                       & 10 & Network Science \\
				MASSACHUSETTS INSTITUTE OF TECHNOLOGY MIT     & 10 & Journal of Complex Networks \\
				\bottomrule
		\end{tabular}}
	\end{table}

	\subsubsection*{Interpretation of Citation and Publication Metrics}
	
	To assess the research impact and scholarly influence of the three selected journals, we computed key bibliometric indicators, summarized in Table \ref{table.tab3}. These metrics provide a structured evaluation of citation patterns, research productivity, and author contributions. Specifically, we analyze the number of cited publications (NCP), the proportion of cited publications (PCP), and the average citations per cited publication (CCP) to assess citation reach. Additionally, we employ widely used impact measures such as the h-index, g-index, and i-index at multiple citation thresholds to capture both depth and breadth of influence within the field. These indicators collectively offer insights into the relative standing and thematic focus of each journal in social and complex network research.
	\begin{table*}[htbp]
		\tiny
		\caption{Citation and Publication Related Metrics}\label{table.tab3}
		\begin{tabularx}{\textwidth}{l|r|r|r|r|r|r|r|c}
			\toprule
			Journal & NCP & PCP & CCP & h-index & g-index & i-10 index & i-100 index & i-200 index \\
			\midrule
			Social Networks & 1462 & 0.939000 & 64.510000 & 126 & 264 & 1004 & 170 & 69 \\
			Network Science & 252 & 0.834400 & 14.680000 & 23 & 52 & 78 & 6 & 1 \\
			Journal of Complex Networks & 431 & 0.786500 & 18.500000 & 34 & 78 & 138 & 10 & 5 \\
			\bottomrule
		\end{tabularx}
	\end{table*}
	\\textit{Social Networks} demonstrates the highest citation influence, with 1,462 cited publications (NCP) and a proportion of cited publications (PCP) of 93.9\%. This indicates a strong citation uptake, reaffirming its status as a leading outlet for social network research \cite{Freeman2004, Borgatti2009}. The average citations per cited publication (CCP = 64.51) is considerably higher than in the other two journals, suggesting that publications in \textit{Social Networks} tend to be more widely cited. The high h-index (126) and g-index (264) further highlight the substantial impact of its articles \cite{moody2004sociology}.
	
	In contrast, \textit{Network Science} and \textit{Journal of Complex Networks}, both more recent journals, exhibit lower citation metrics. \textit{Network Science} has a PCP of 83.4\% and a CCP of 14.68, indicating that while most of its articles receive citations, they accumulate fewer per publication. Similarly, \textit{Journal of Complex Networks} reports a PCP of 78.7\% and a CCP of 18.5, reflecting its role in a more specialized computational and theoretical niche \cite{Newman2010, Barabasi2016}.
	
	The distribution of highly cited publications aligns with these patterns. \textit{Social Networks} has 1,004 papers exceeding 10 citations (i-10), with 170 surpassing 100 citations (i-100) and 69 exceeding 200 citations (i-200). This far surpasses the counts in \textit{Network Science} (i-100 = 6, i-200 = 1) and \textit{Journal of Complex Networks} (i-100 = 10, i-200 = 5), reinforcing its dominant position in the field \cite{wasserman1994social, Watts1998}.
	
	Overall, these metrics reflect the distinct roles of each journal: \textit{Social Networks} as the primary venue for empirical and applied social network studies, \textit{Journal of Complex Networks} as a hub for mathematical and computational approaches, and \textit{Network Science} as an interdisciplinary bridge \cite{Barabasi2009}. The citation trends highlight the field’s evolution, with computational and theoretical research gaining visibility while traditional empirical studies maintain the highest impact.

\section{Science Mapping} 

Science mapping provides a visual and structural representation of the intellectual landscape of a research domain \cite{Cobo2011, Börner2010}. By analyzing co-citation networks, bibliographic coupling, and keyword co-occurrence patterns, this approach uncovers thematic clusters, influential works, and evolving research trends \cite{Aria2017, Chen2006}. 

In this study, we employ three widely used bibliometric tools—VOSviewer, Gephi, and CiteSpace—to conduct science mapping. VOSviewer is utilized for constructing and visualizing bibliometric networks \cite{vanEck2010vosviewer}, Gephi enables advanced network analysis of scholarly collaboration \cite{bastian2009gephi}, and CiteSpace facilitates the identification of emerging trends and intellectual turning points \cite{Chen2017}. Through these methods, we systematically examine the thematic development, conceptual linkages, and structural properties of the selected journals within the field of social and complex network research.

\subsection{Citation Analysis}

For all datasets, we use the \textit{g-index} to regulate the network size. The \textit{g-index} is defined by the equation:

$$g^2 \leq k \sum_{i \leq g} c_i, \quad k \in \mathbb{Z}^+$$,

where we set $k = 25$ to constrain the network size.

Citation analysis identifies the most influential publications within a research domain by examining citation counts and scholarly impact \cite{Garfield1972, Small1973}. Highly cited papers shape the intellectual structure of a field, serving as foundational works that influence subsequent research \cite{Leydesdorff1998, Moed2005}. 

Table \ref{table.tab4} presents the top 10 most cited papers in \textit{Social Networks}, highlighting key contributions to network analysis. Freeman’s (1979) seminal work on centrality remains the most influential, shaping decades of research on network metrics \cite{Freeman1979}. Other highly cited studies introduce methodological advancements such as weighted centrality measures \cite{Opsahl2010}, stochastic blockmodels \cite{Holland1983}, and exponential random graph models \cite{Robins2007}. The dominance of works on centrality, community structure, and network modeling reflects the journal’s strong emphasis on both empirical applications and theoretical innovations.

These citation patterns underscore \textit{Social Networks} as a critical venue for advancing methodologies in social network analysis. Further analysis of co-citation networks will reveal how these studies interact and influence emerging research trajectories.

	\subsubsection*{Social Networks}
	
	The citation network of \textit{Social Networks} exhibits a centralized hub structure, where key publications serve as bridges across different research areas. These influential works facilitate knowledge diffusion, linking empirical, computational, and theoretical approaches. To further investigate their impact, we analyze degree centrality, betweenness centrality, and modularity, identifying emerging research trends and intellectual shifts \cite{snijders2010stochastic, Holland1983}.
	
	\begin{table*}[htbp]
		\centering
		\scriptsize
		\caption{Top 10 Most Cited Papers in \textit{Social Networks}}
		\label{table.tab4}
		\begin{tabularx}{\textwidth}{llrl}
			\toprule
			\textbf{Author(s)} & \textbf{Title} & \textbf{Year} & \textbf{Citations} \\
			\midrule
			Freeman, LC & Centrality in Social Networks: Conceptual Clarification & 1979 & 10811 \\
			Opsahl, T & Node Centrality in Weighted Networks: Generalizing Degree and Shortest Path & 2010 & 2269 \\
			Borgatti, SP & Centrality and Network Flow & 2005 & 2108 \\
			Adamic, LA & Friends and Neighbors on the Web & 2003 & 1786 \\
			Holland, PW & Stochastic Blockmodels: First Steps & 1983 & 1690 \\
			Newman, MEJ & A Measure of Betweenness Centrality Based on Random Walks & 2005 & 1420 \\
			Snijders, TAB & Introduction to Stochastic Actor-Based Models for Network Dynamics & 2010 & 1312 \\
			Borgatti, SP & Models of Core/Periphery Structures & 1999 & 1282 \\
			Robins, G & An Introduction to Exponential Random Graph (\textit{p*}) Models & 2007 & 1227 \\
			Seidman, SB & Network Structure and Minimum Degree & 1983 & 1184 \\
			\bottomrule
		\end{tabularx}
	\end{table*}
	
\begin{figure}[htbp]
		\centering
		\includegraphics[width=\columnwidth]{"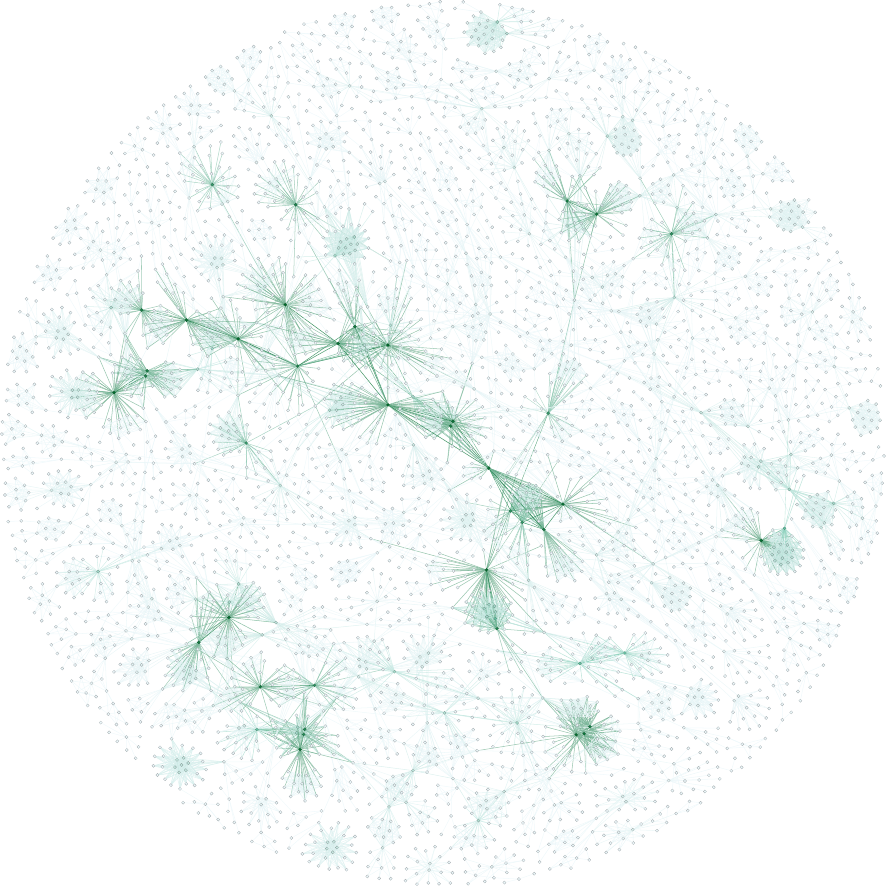"}
		\caption{Largest Connected Component in \textit{Social Networks}}
		\label{fig.fig3}
	\end{figure}

	Figure \ref{fig.fig3} visualizes the largest connected component of the citation network in \textit{Social Networks}. Darker nodes with high connectivity represent foundational works that anchor major research clusters. These dense structures correspond to specialized subfields with strong internal citations, while lighter, more dispersed nodes indicate peripheral contributions.
	
\begin{figure}[htbp]
		\centering
		\includegraphics[width=\columnwidth]{"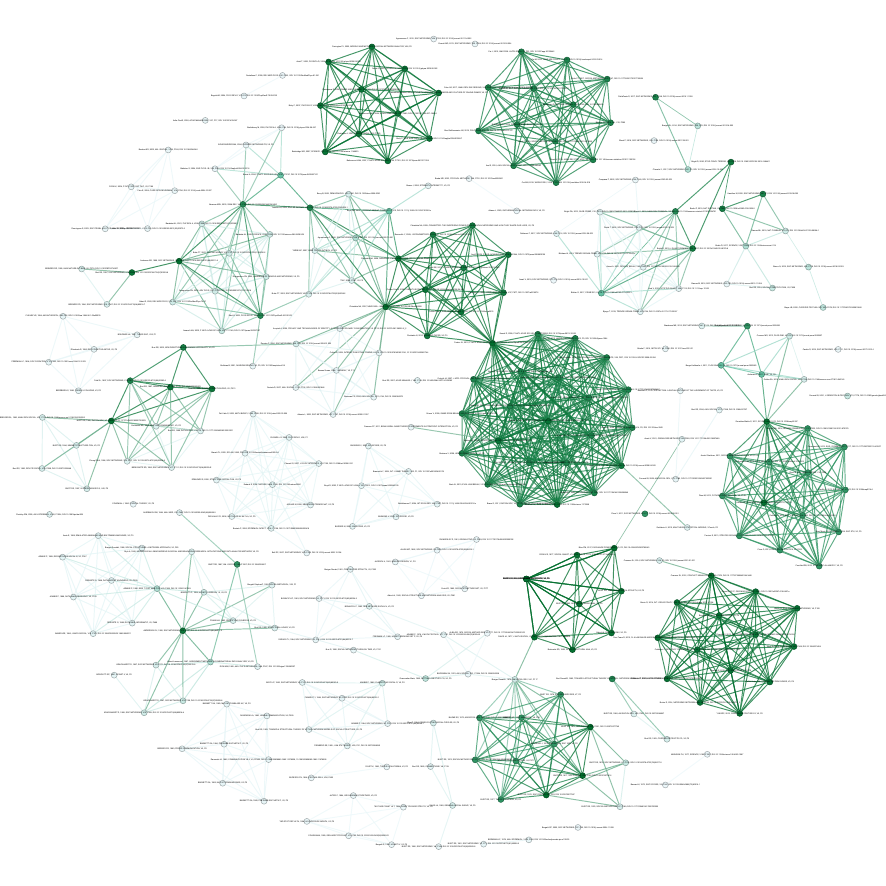"}
		\caption{High-Degree Clusters in \textit{Social Networks}}
		\label{fig.fig4}
	\end{figure}

	Further examination of citation patterns reveals that high-degree nodes form tightly connected clusters, suggesting strong citation reciprocity within subfields. This is illustrated in Figure \ref{fig.fig4}, where darker colors indicate nodes with higher degrees. The network layout follows the Fruchterman-Reingold algorithm \cite{fruchterman1991graph}, with the degree range set between 21 and 61 to highlight structural patterns.
	
	\subsubsection*{Network Science}

	The citation network of \textit{Network Science} in Table \ref{table.tab5} reveals a distinct intellectual structure, reflecting its role in integrating interdisciplinary research on complex networks. High-impact papers in this journal span theoretical advancements, computational methodologies, and empirical applications. The network is characterized by key publications that serve as foundational references in network analysis, algorithmic development, and data-driven modeling.
	
	\begin{table*}[htbp]
		\centering
		\scriptsize
		\caption{Top 10 Most Cited Papers in \textit{Network Science}}
		\label{table.tab5}
		\begin{tabularx}{\textwidth}{llrl}
			\toprule
			\textbf{Author(s)} & \textbf{Title} & \textbf{Year} & \textbf{Citations} \\
			\midrule
			Barabási, AL  & Network Science  & 2016  & 901 \\
			 Barabási, AL  & Personal Introduction  & 2016  & 191 \\
			Génois, M  & Data on Face-to-Face Contacts in an Office Building Suggest a Low-Cost  & 2015  & 147 \\
			 Bothorel, C  & Clustering Attributed Graphs: Models, Measures, and Methods  & 2015  & 136 \\
			Staudt, CL  & NetworKit: A Tool Suite for Large-Scale Complex Network Analysis  & 2016  & 121 \\
			Block, P  & Multidimensional Homophily in Friendship Networks  & 2014  & 115 \\
			Rodriguez, MG  & Uncovering the Structure and Temporal Dynamics of Information  & 2014  & 87 \\
			Neal, ZP  & How Small is it? Comparing Indices of Small Worldliness  & 2017  & 45 \\
			 Leifeld, P  & A Theoretical and Empirical Comparison of the Temporal Exponential  & 2019  & 43 \\
			Elmer, T  & The Co-Evolution of Emotional Well-Being with Weak and Strong Friendship  & 2017  & 41 \\
			\bottomrule
		\end{tabularx}
	\end{table*}
	
\begin{figure}[htbp]
		\centering
		\includegraphics[width=\columnwidth]{"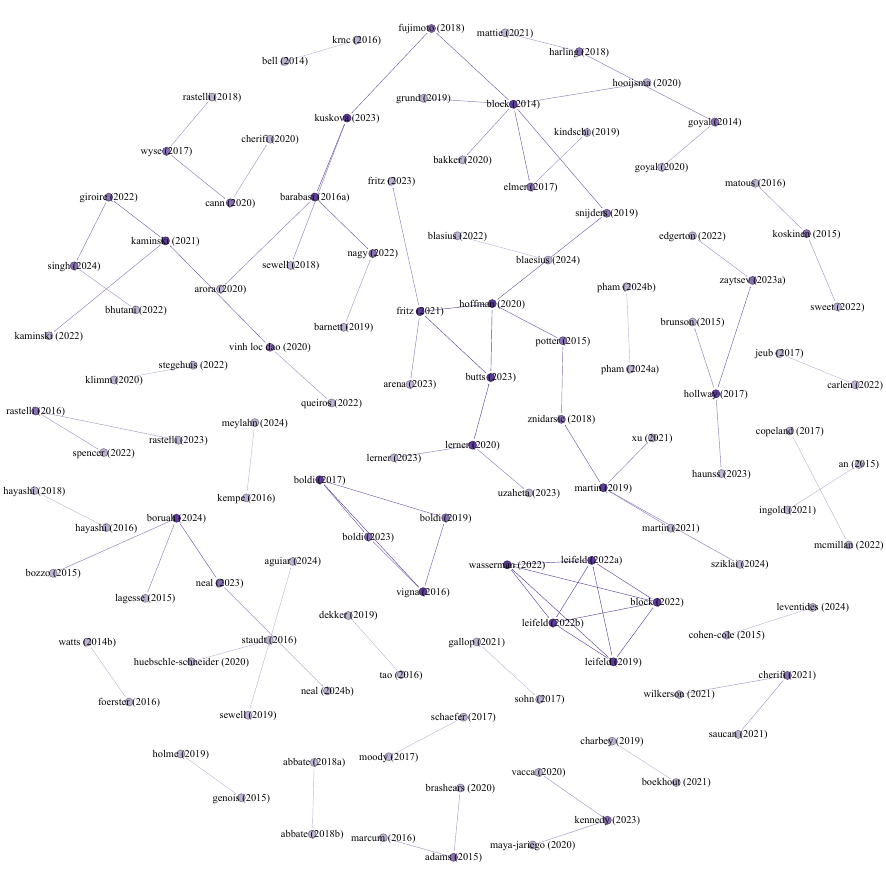"}
		\caption{Highly Cited Papers in \textit{Network Science}}
		\label{fig.fig5}
	\end{figure}

	The citation network of \textit{Network Science} shown in Figure \ref{fig.fig5} exhibits a more fragmented structure compared to \textit{Social Networks}, suggesting that authors in this journal engage in relatively independent research trajectories with fewer direct collaborations. The distribution of citations is more dispersed, indicating that influential works are spread across multiple subfields rather than concentrated in a few key publications.
	
	\subsubsection*{Journal of Complex Networks}
	
	The \textit{Journal of Complex Networks} (JCN) exhibits \textbf{moderate connectivity} in its citation and collaboration structures, a characteristic supported by previous studies on academic network structures \cite{newman2001scientific, Liu2005}. Compared to \textit{Network Science} (NS), JCN demonstrates slightly better overall connectivity, suggesting a higher degree of scholarly interaction \cite{Barabasi2002}. However, its \textbf{largest connected component is significantly smaller than that of Social Networks (SN)}, indicating a more fragmented intellectual structure \cite{Otte2002, Borgatti2009}.

	\begin{figure}[htbp]
		\centering
		\includegraphics[width=\columnwidth]{"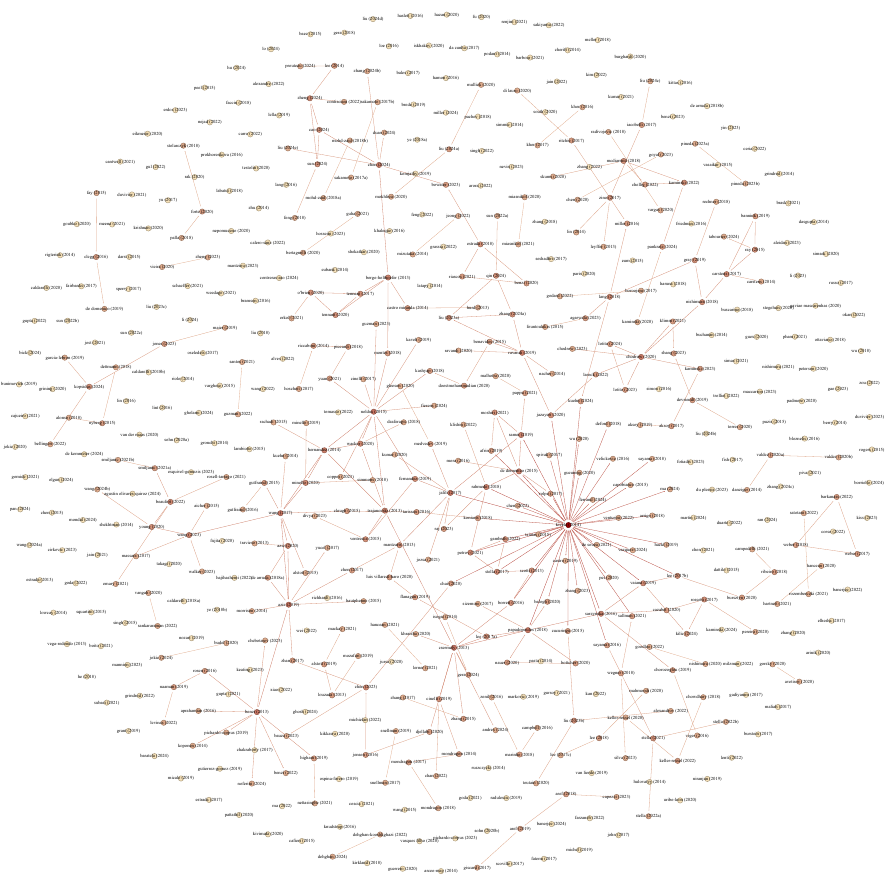"}
		\caption{Highly cited papers in \textit{Journal of Complex Networks}}
		\label{fig.fig6}
	\end{figure}
	
	JCN's co-authorship network reveals a \textbf{dispersed collaboration pattern} in Figure \ref{fig.fig6}, with relatively independent research teams contributing to distinct methodological subfields. Unlike \textit{Social Networks}, which features dense collaborative clusters, JCN primarily consists of \textbf{smaller, loosely connected research groups}, reflecting its emphasis on specialized theoretical and computational approaches rather than broad empirical studies.
	
	These structural characteristics suggest that while JCN supports a diverse range of research topics in complex networks, its interdisciplinary integration remains limited. The relatively low level of cross-group citation and collaboration indicates that JCN could benefit from fostering \textbf{stronger interdisciplinary engagement}, potentially enhancing the coherence and impact of its scholarly network.
	
	\begin{figure}[htbp]
		\centering
		\includegraphics[width=\columnwidth]{"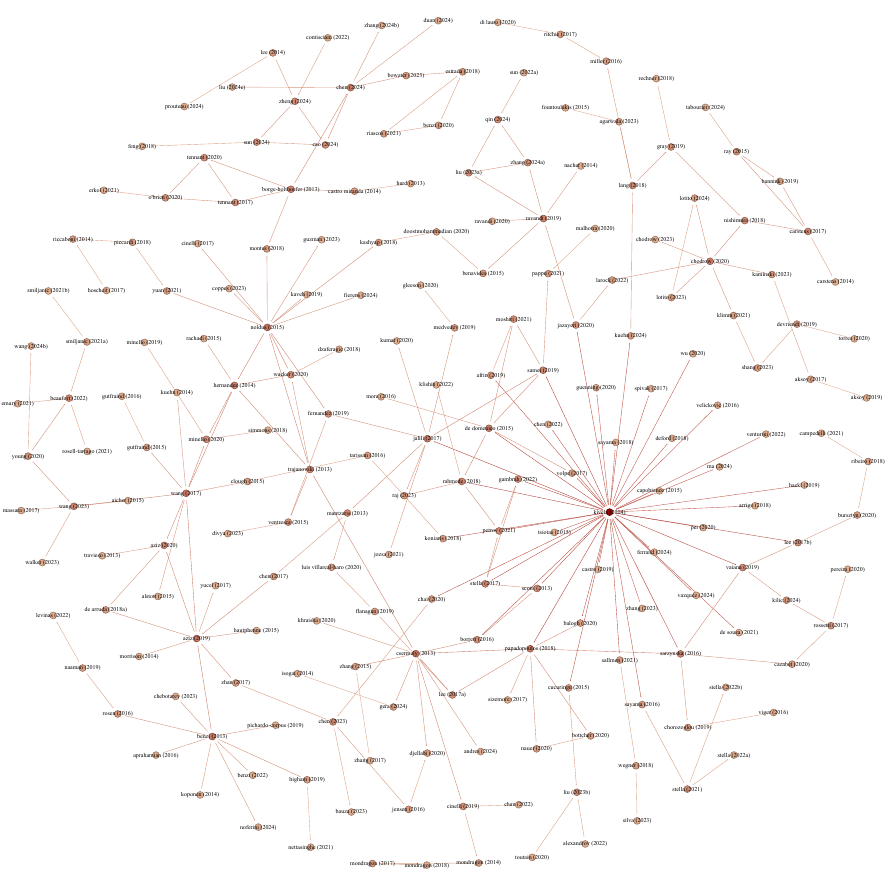"}
		\caption{Largest connected component in \textit{Journal of Complex Networks}.}
		\label{fig.fig7}
	\end{figure}
	
	We also observe that while JCN exhibits higher connectivity than \textit{Network Science}, its citation structure remains \textbf{fragmented}, with relatively few high-degree nodes anchoring the network. This is evident in Figure \ref{fig.fig6}, where darker colors represent higher-degree nodes, highlighting influential papers within the network.
	
	\subsection{Co-citation Analysis}
	
	Co-citation analysis examines the intellectual structure of a research field by identifying relationships between publications that are frequently cited together \cite{Small1973}. In this section, we conduct a co-citation analysis across the three selected journals—\textit{Social Networks}, \textit{Journal of Complex Networks}, and \textit{Network Science}—to uncover key scholarly influences and thematic clusters within network science.
	
	By mapping co-citation patterns, we aim to:
	\begin{itemize}
		\item Identify foundational and highly co-cited works that shape the field.
		\item Detect thematic clusters that highlight dominant research areas.
		\item Examine the evolution of intellectual discourse across the three journals.
	\end{itemize}
	
	The analysis is conducted using \textbf{VOSviewer, CiteSpace, and Gephi}\cite{vanEck2010vosviewer,Chen2006,bastian2009gephi}, employing network visualization techniques to reveal structural patterns. The following sections present the results for each journal, highlighting the most influential papers and their co-citation relationships. 
	
	\subsubsection*{Social Networks}
	
	To examine the intellectual foundations of research published in \textit{Social Networks}, we conducted a co-citation analysis using the 40,748 references cited across the dataset. We applied a minimum co-citation threshold of 10, resulting in 789 publications that met the criterion. These publications formed six distinct research clusters, indicating core thematic areas within the field.
	
\begin{figure}[htbp]
		\centering
		\includegraphics[width=\columnwidth]{"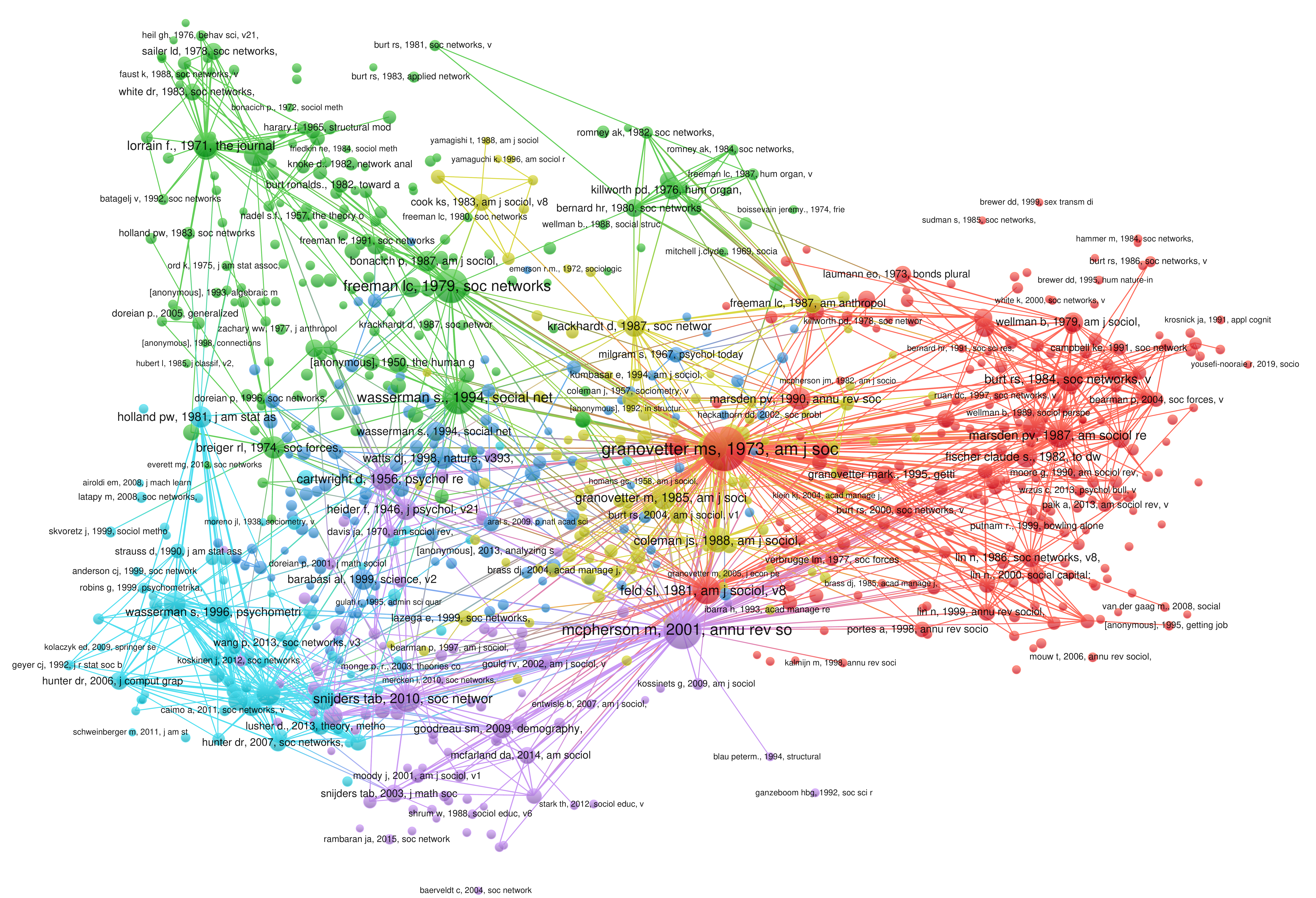"}
		\caption{Co-citation network of \textit{Social Networks}}
		\label{fig.fig8}
	\end{figure}

	Figure \ref{fig.fig8} visualizes the co-citation network, where node size represents citation frequency, and edge thickness denotes co-citation strength. The modular structure reveals highly cited seminal works at the core of each cluster, surrounded by peripheral contributions that extend or apply foundational theories. 
	
	The six detected clusters represent distinct intellectual traditions within social network research:
	
	\begin{itemize}
		\item \textbf{Structural Network Analysis}: This cluster centers around foundational works on network centrality, small-world networks, and structural balance, with key contributions from Freeman (1979)~\cite{Freeman1979}, Borgatti \& Everett (1999)~\cite{borgatti1999core}, and Watts \& Strogatz (1998)~\cite{Watts1998}.
		\item \textbf{Statistical Network Models}: Influenced by exponential random graph models (ERGMs) and stochastic blockmodels, this cluster includes seminal works by Holland \& Leinhardt (1981)~\cite{holland1981exponential} and Snijders (2001)~\cite{snijders2001statistical}.
		\item \textbf{Computational and Algorithmic Methods}: Emerging as an increasingly influential area, this cluster includes research on network inference, community detection, and machine learning applications in network science.
		\item \textbf{Social Capital and Influence Diffusion}: Studies on social influence, information diffusion, and network externalities form another major research stream, building on Granovetter (1973)~\cite{granovetter1973strength} and Burt (1992)~\cite{Burt1992}.
		\item \textbf{Longitudinal and Dynamic Networks}: This cluster focuses on temporal network analysis, including actor-oriented models for evolving networks (e.g., Snijders et al., 2010)~\cite{snijders2010stochastic}.
		\item \textbf{Applications in Social and Policy Sciences}: The final cluster captures interdisciplinary applications of social network analysis in areas such as epidemiology, political networks, and organizational studies.
	\end{itemize}

	These clusters highlight the dual nature of the field, where classical theoretical frameworks continue to shape research trajectories while computational advances and interdisciplinary applications drive emerging themes. The presence of well-defined yet interconnected clusters suggests that social network research maintains a balance between theoretical consolidation and methodological diversification.
	
	\subsubsection*{Network Science}
		
		A co-citation analysis was also performed to identify the foundational literature shaping research published in \textit{Network Science}. The dataset contained 10,932 cited references, of which only 59 met the minimum co-citation threshold of 10. These references were grouped into three distinct clusters, representing key intellectual domains within the journal.
		
		\begin{figure}[htbp]
			\centering
			\includegraphics[width=\columnwidth]{"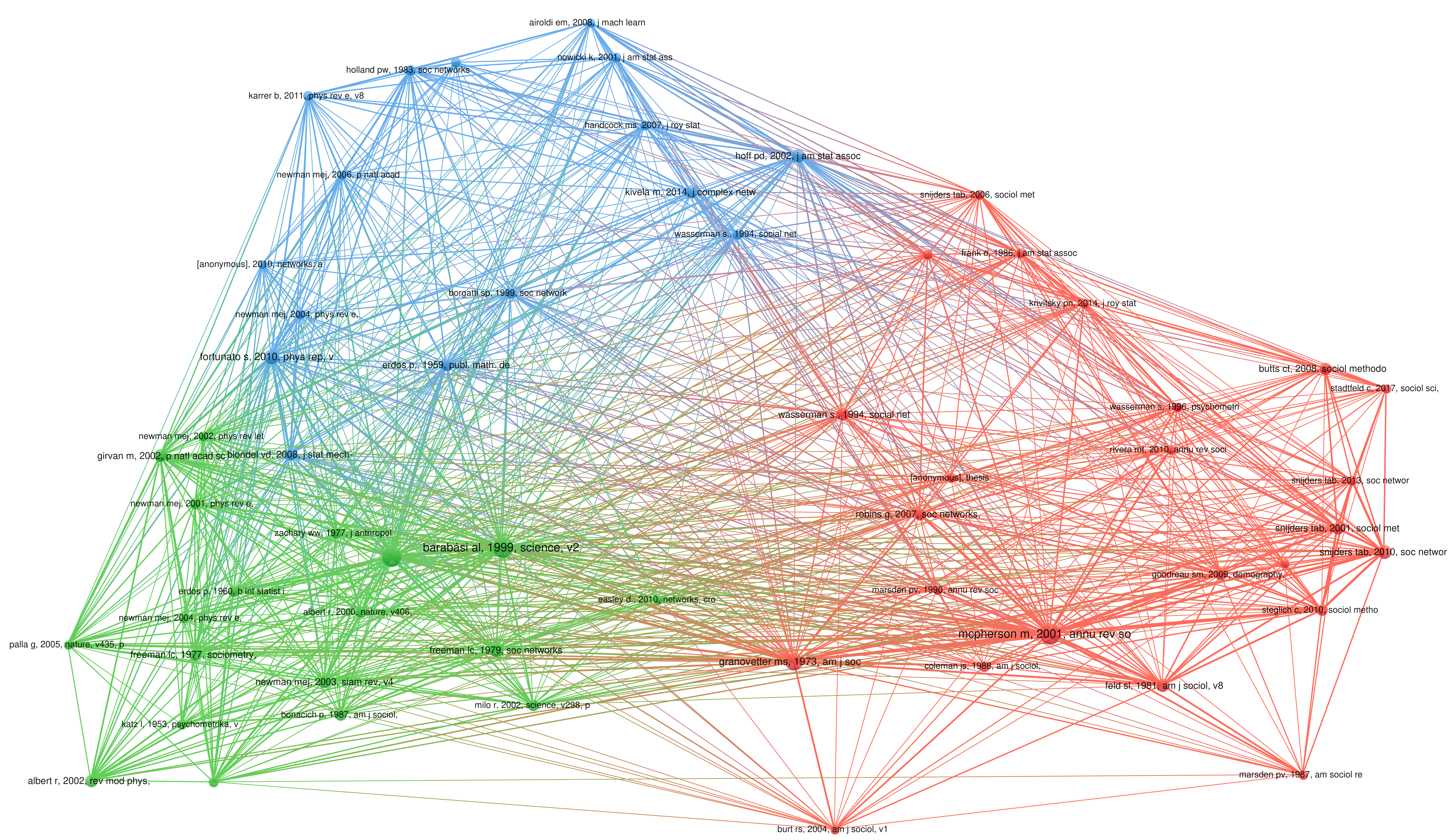"}
			\caption{Co-citation network of \textit{Network Science}.}
			\label{fig.fig9}
		\end{figure}
		
		Figure \ref{fig.fig9} presents the resulting co-citation network, illustrating how frequently cited works are interconnected through shared scholarly influence. The structure of this network reveals a core set of highly referenced publications that serve as conceptual anchors, linking diverse areas of inquiry within network science. The clusters indicate distinct thematic orientations, with some references contributing to theoretical advancements in network modeling, while others emphasize computational techniques or interdisciplinary applications.
		
		The limited number of publications surpassing the co-citation threshold suggests that research in \textit{Network Science} builds upon a relatively concentrated set of seminal works. This reflects the journal’s specialized focus, where foundational theories and methodological innovations form the backbone of scholarly discourse. The presence of clear clusters also highlights disciplinary coherence, while inter-cluster linkages suggest ongoing cross-pollination between different research approaches.

		\subsubsection*{Journal of Complex Networks}
		
		To analyze the foundational literature shaping research in \textit{Journal of Complex Networks}, we conducted a co-citation analysis based on 16,758 cited references. Applying a minimum co-citation threshold of 10, we identified 136 publications that met this criterion. These references formed five distinct research clusters, each representing a key intellectual domain within complex network studies.
		
		\begin{figure}[htbp]
			\centering
			\includegraphics[width=\columnwidth]{"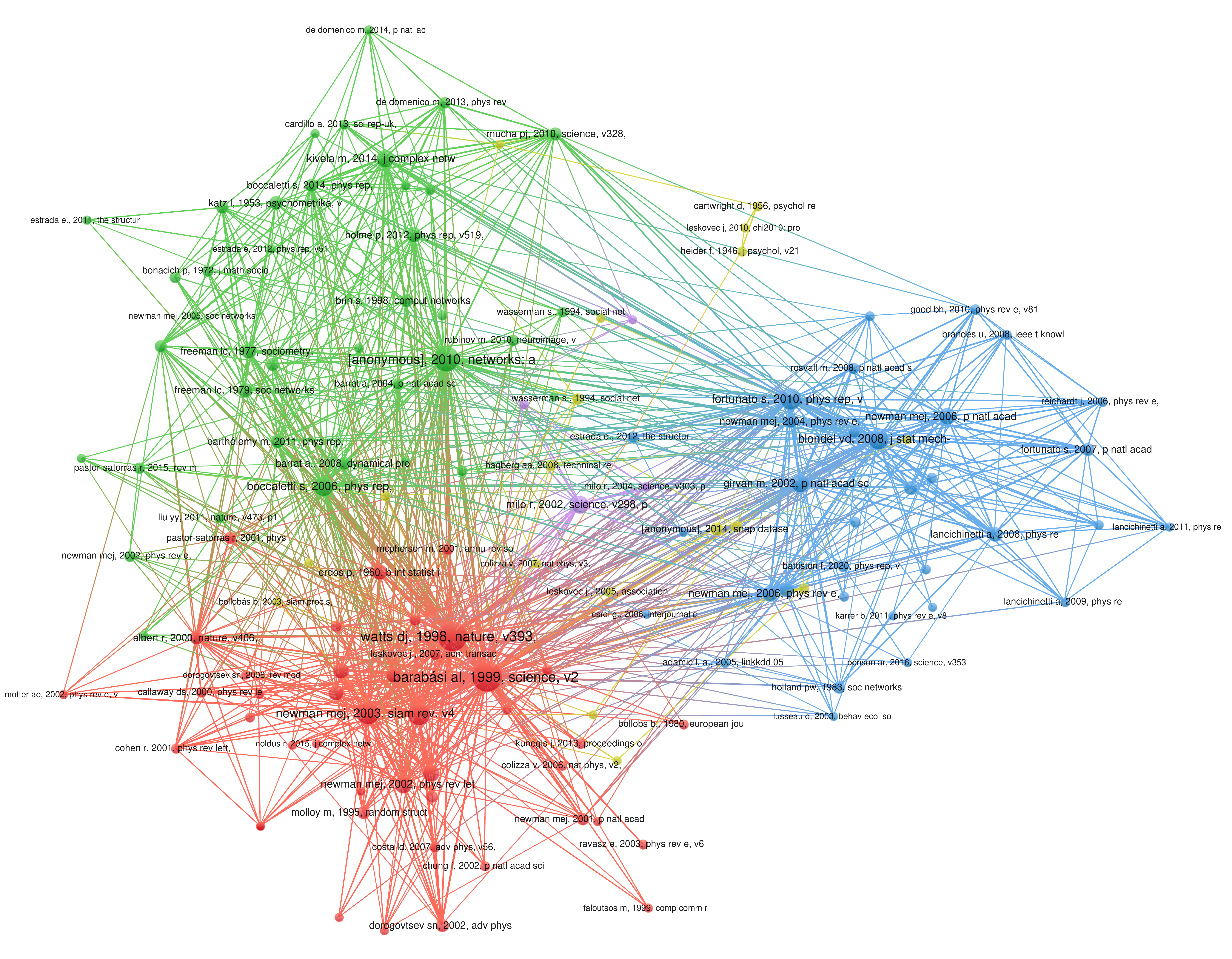"}
			\caption{Co-citation network of \textit{Journal of Complex Networks}.}
			\label{fig.fig10}
		\end{figure}
		
		The co-citation network, visualized in Figure \ref{fig.fig10}, reveals the structural organization of highly cited references in JCN. Prominent works in network science, such as those by Newman, Barabási, and Watts, emerge as central nodes, highlighting their widespread influence across multiple research clusters. The presence of studies on modularity detection, network dynamics, and multilayer networks suggests that JCN is deeply engaged with theoretical and methodological advancements in complex systems.
		
		The five identified clusters indicate thematic subdivisions within the field:
		\begin{itemize}
			\item \textbf{Network Structure and Dynamics:} Fundamental studies on network topology, small-world effects, and preferential attachment.
			\item \textbf{Community Detection and Modularity:} Key algorithms for identifying structural patterns in complex networks.
			\item \textbf{Multilayer and Temporal Networks:} Research focusing on network evolution and interactions across different layers.
			\item \textbf{Statistical Mechanics of Networks:} Theoretical frameworks derived from physics-based approaches.
			\item \textbf{Applications in Biology and Social Systems:} Studies applying network models to real-world datasets, particularly in epidemiology and information diffusion.
		\end{itemize}
		
		Overall, JCN’s co-citation structure underscores its focus on methodological innovation and theoretical contributions to complex network analysis. The dispersion of research clusters suggests a broad yet specialized research landscape, where computational approaches are deeply integrated with applied network studies.
		
		The co-citation analysis across the three journals—\textit{Social Networks}, \textit{Network Science}, and \textit{Journal of Complex Networks}—reveals distinct intellectual structures and thematic orientations within the broader domain of network research.
		
		\begin{itemize}
			\item \textbf{Social Networks (SN):} Exhibits a well-established and diversified co-citation network, with six thematic clusters encompassing classical structural analysis, statistical modeling, computational techniques, and applied social sciences. The breadth of co-cited works reflects the journal’s long-standing role in advancing both theoretical and empirical social network research.
			
			\item \textbf{Network Science (NS):} Features a more compact intellectual structure, with only three co-citation clusters. This indicates a concentrated reliance on foundational network theories and computational methods. The journal’s specialized nature suggests that its research community is strongly anchored in a core set of influential studies.
			
			\item \textbf{Journal of Complex Networks (JCN):} Presents a moderately interconnected co-citation network with five clusters, reflecting its emphasis on mathematical modeling, community detection, multilayer networks, and statistical mechanics. The presence of applied research in biological and social systems underscores its interdisciplinary scope, bridging theoretical and practical applications.
			
		\end{itemize}
		
		These findings highlight both the convergence and divergence of intellectual influences across the three journals. \textit{Social Networks} maintains the most extensive and heterogeneous co-citation landscape, integrating diverse methodological traditions. \textit{Network Science} exhibits a concentrated scholarly foundation, reinforcing its role as a hub for theoretical and computational advances. \textit{Journal of Complex Networks} occupies an intermediary position, blending formal network analysis with real-world applications.
		
		The observed structural differences suggest that while network research continues to evolve, disciplinary specialization persists. Future bibliometric analyses could further investigate citation flows between these journals, identifying cross-fertilization of ideas and emerging interdisciplinary trends.
		
		\subsection{Co-word Analysis}
		
		Co-word analysis examines the conceptual structure of a research field by identifying patterns of keyword co-occurrence in scholarly publications. By mapping the relationships between frequently co-occurring terms, this method reveals dominant research themes, emerging trends, and the evolution of discourse within a field~\cite{callon1983co, coulter1998co, ding2001co}.

		In this study, we conduct a co-word analysis across \textit{Social Networks}, \textit{Network Science}, and \textit{Journal of Complex Networks} to:
		
		\begin{itemize}
			\item Identify high-frequency keywords that define core topics within each journal.
			\item Detect thematic clusters and conceptual linkages between research areas.
			\item Explore temporal shifts in keyword usage to track emerging trends.
		\end{itemize}
		
		The analysis is performed using \textbf{VOSviewer} and \textbf{CiteSpace}, leveraging network visualization techniques to illustrate keyword co-occurrence structures. The resulting networks provide insight into how key concepts interact and evolve across different subfields of network science.
		
		The following sections present the results for each journal, highlighting major thematic clusters and their implications for the intellectual development of the field.
		
		\begin{figure*}
			\centering
			\caption{\label{fig.fig11}Density Visulization}
			\begin{minipage}[b]{0.32\textwidth}
				\centering
				\includegraphics[width=\linewidth]{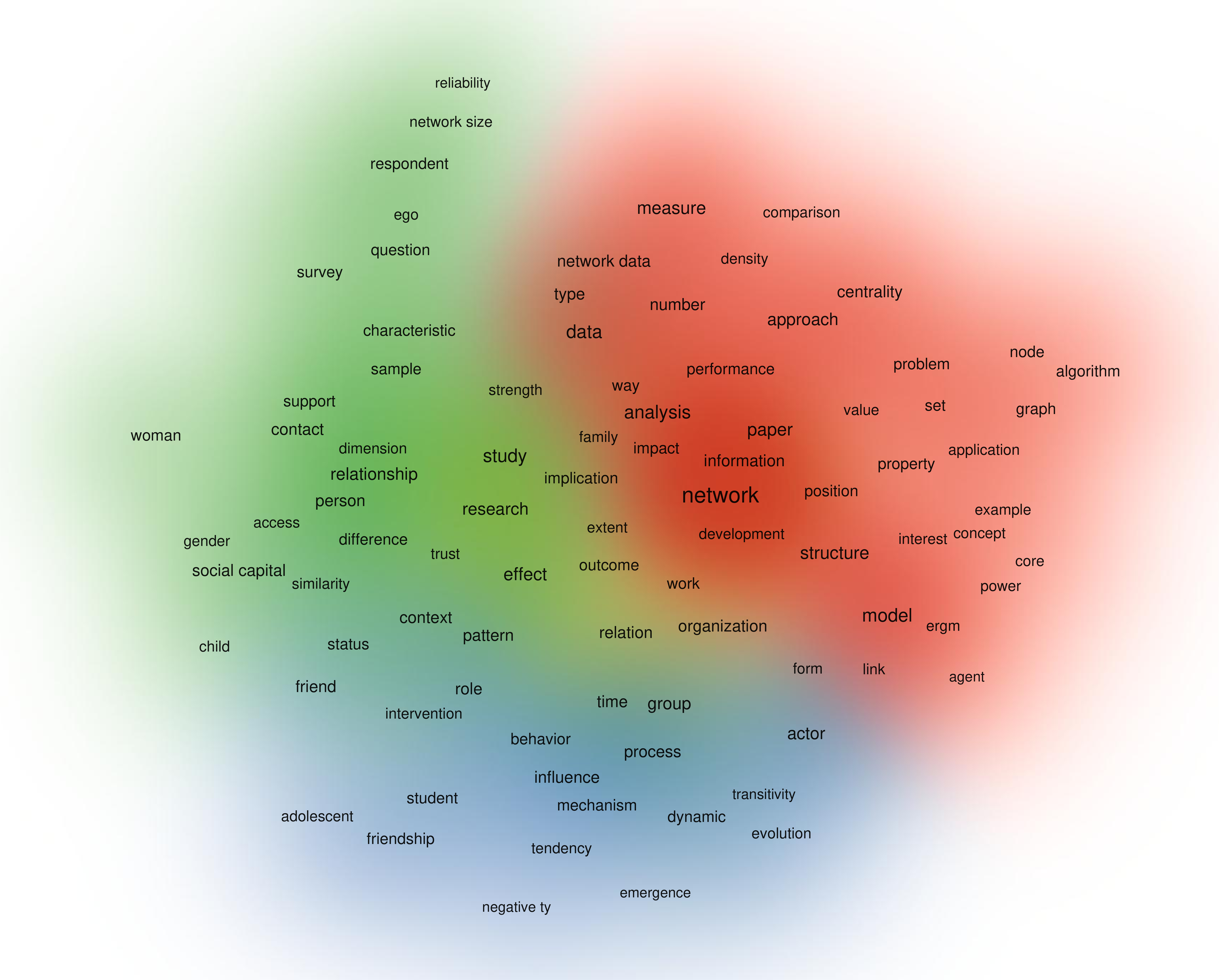}
				(a) Social Networks Density Visulization
			\end{minipage}
			\begin{minipage}[b]{0.32\textwidth}
				\centering
				\includegraphics[width=\linewidth]{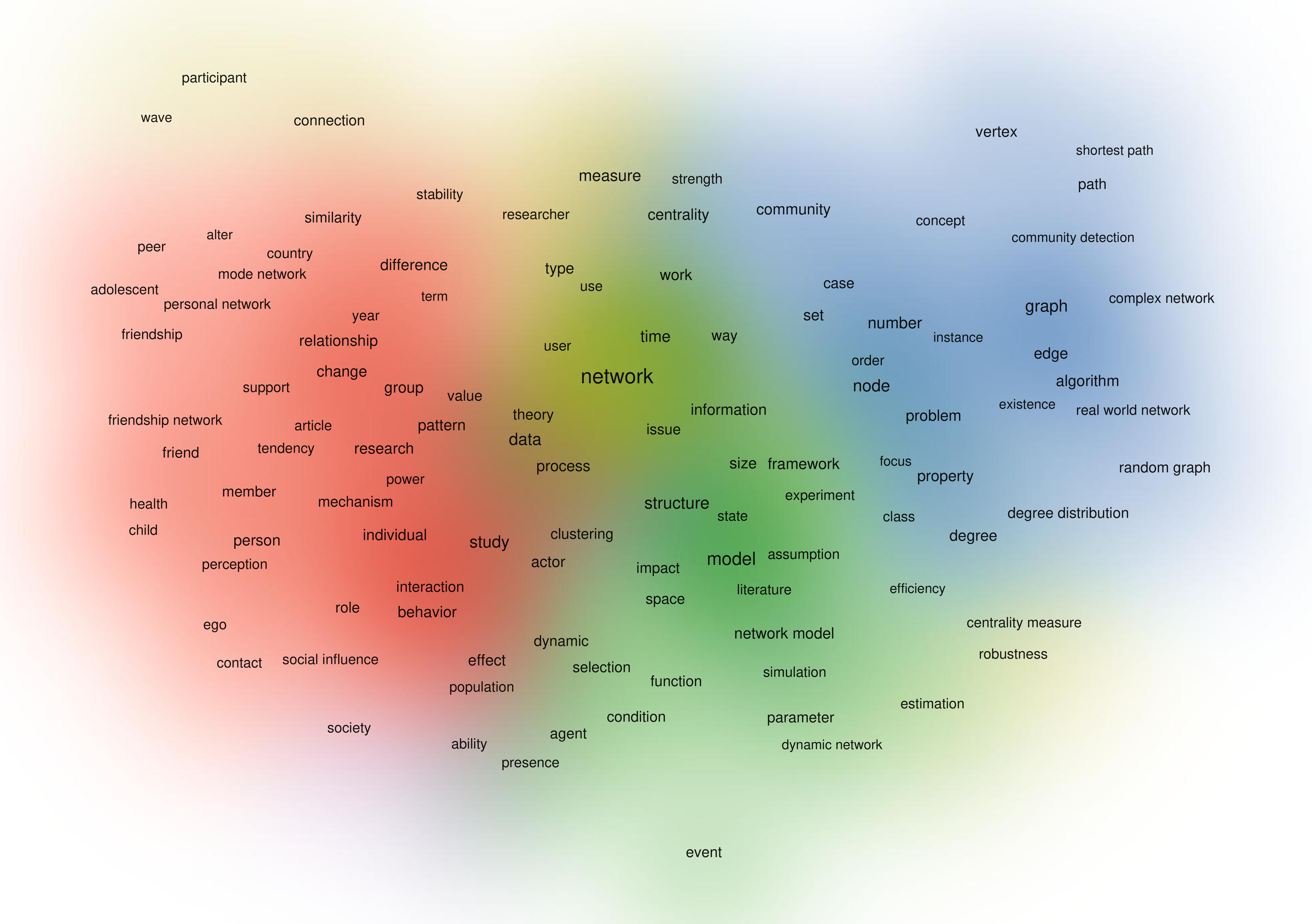}
				(b) Network Science Density Visulization
			\end{minipage}
			\begin{minipage}[b]{0.32\textwidth}
				\centering
				\includegraphics[width=\linewidth]{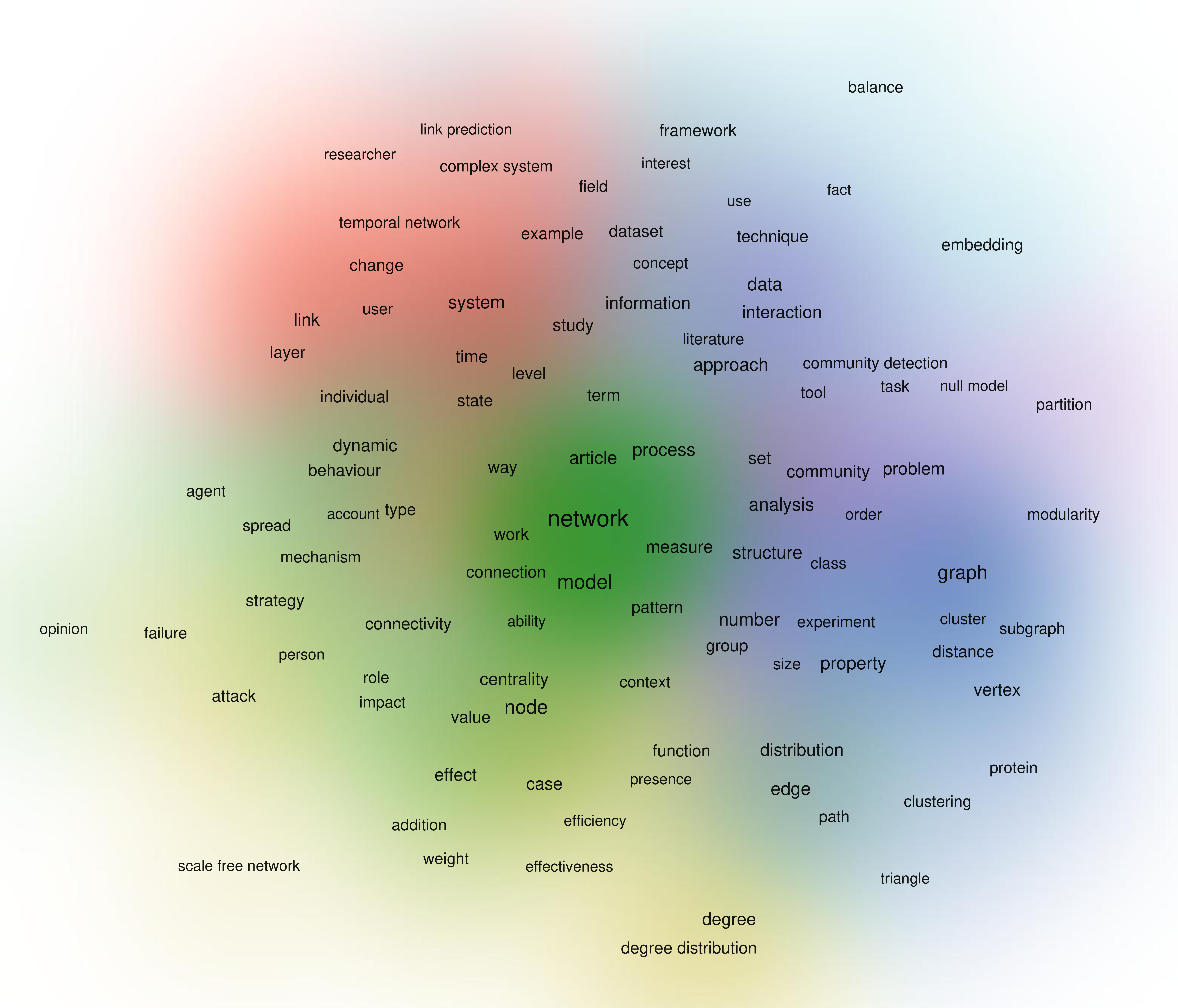}
				(c) Journal of Complex Networks Density Visulization
			\end{minipage}
		\end{figure*}
		
		The density visualization of co-word networks in Figure \ref{fig.fig11} provides insights into the conceptual structures of research published in \textit{Social Networks}, \textit{Journal of Complex Networks}, and \textit{Network Science}. By comparing their density layouts, we observe key differences in the thematic concentration and distribution of research topics across these journals.
		
		\begin{itemize}
			\item \textbf{Social Networks}: The density map of \textit{Social Networks} exhibits a well-defined structure with multiple highly concentrated regions. Core research themes such as social capital, centrality, and network structure appear densely clustered, indicating a long-established and cohesive intellectual foundation. The strong concentration of keywords suggests a mature research field with well-integrated concepts.
			
			\item \textbf{Journal of Complex Networks}: The density map of \textit{Journal of Complex Networks} displays a more dispersed pattern with several moderately dense clusters. Topics related to algorithmic approaches, network modeling, and complex systems are prevalent. The distribution suggests a field that, while specialized, is still evolving, with diverse research directions coexisting rather than converging around a few dominant themes.
			
			\item \textbf{Network Science}: The density map for \textit{Network Science} is relatively sparse, with fewer highly concentrated areas. This indicates a more fragmented research landscape, reflecting the journal’s interdisciplinary nature. Unlike \textit{Social Networks}, which has a tightly integrated body of research, \textit{Network Science} includes diverse but loosely connected research themes from physics, computer science, and sociology, resulting in a more diffused density layout.
		\end{itemize}
		
		These differences reflect the distinct epistemic communities served by each journal. \textit{Social Networks} maintains a traditional focus on social theory and empirical analysis, leading to a high-density conceptual core. \textit{Journal of Complex Networks} balances theoretical and computational perspectives, resulting in a moderately cohesive structure. \textit{Network Science} embraces a broad interdisciplinary scope, leading to a lower-density conceptual space with more loosely connected research topics.
		
		\subsection{Co-authorship Analysis}
		
		\subsubsection*{Social Networks}
		
\begin{figure}[htbp]
			\centering
			\includegraphics[width=\columnwidth]{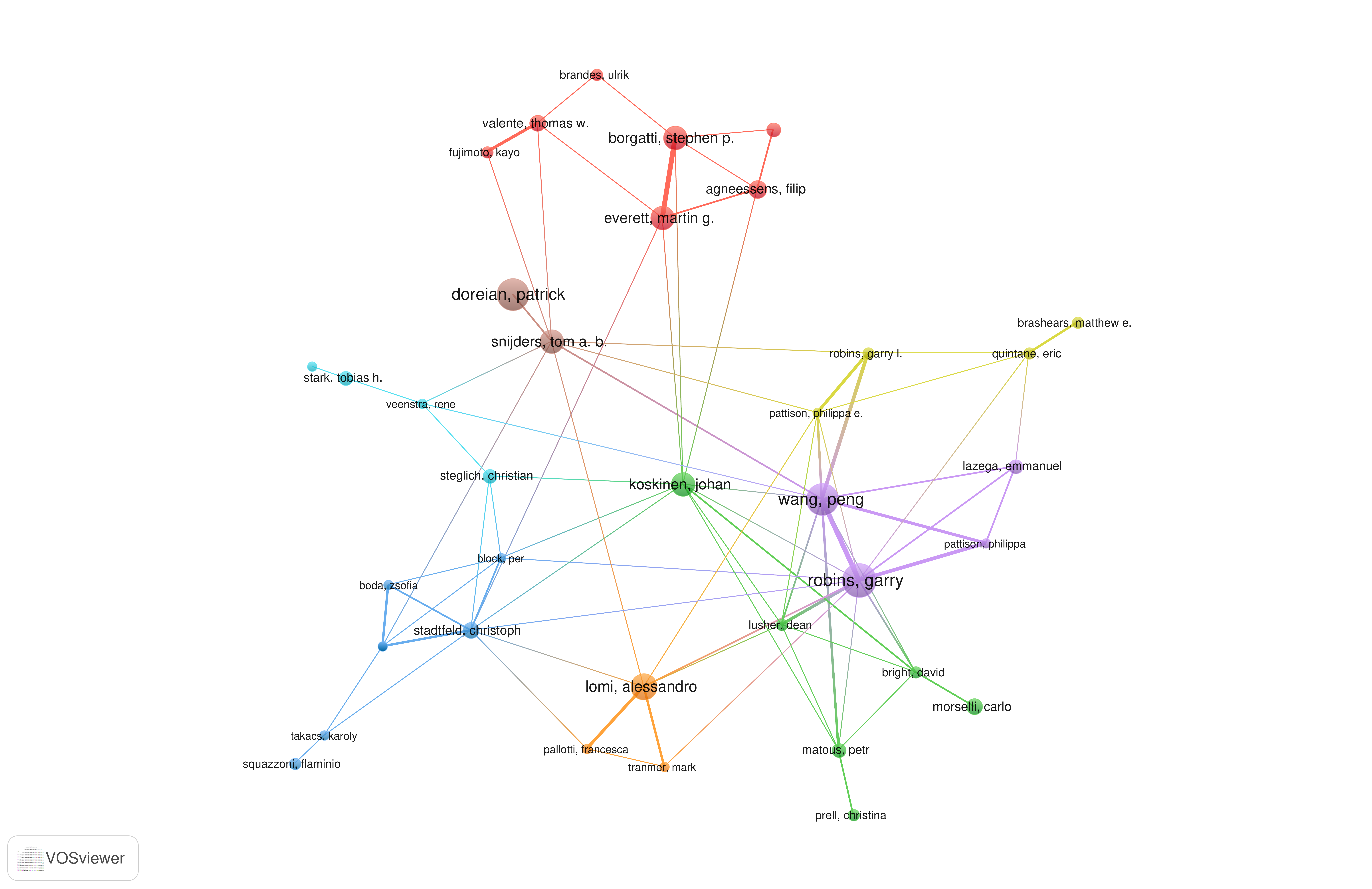}
			\caption{Co-authorship network visualization}
			\label{fig.fig13}
		\end{figure}

		To analyze the co-authorship patterns in the journal \textit{Social Networks}, we constructed a co-authorship network using bibliometric data. The visualization was generated using VOSviewer 1.6.20~\cite{van2010visualizing}, which clusters authors based on their collaborative relationships. Figure \ref{fig.fig13} illustrates the co-authorship network, where nodes represent individual authors, and edges denote co-authorship links. The size of each node corresponds to the author's publication volume, while edge thickness reflects the frequency of co-authorship~\cite{glanzel2004handbook, newman2004coauthorship}.

\begin{figure}[htbp]
			\centering
			\includegraphics[width=\columnwidth]{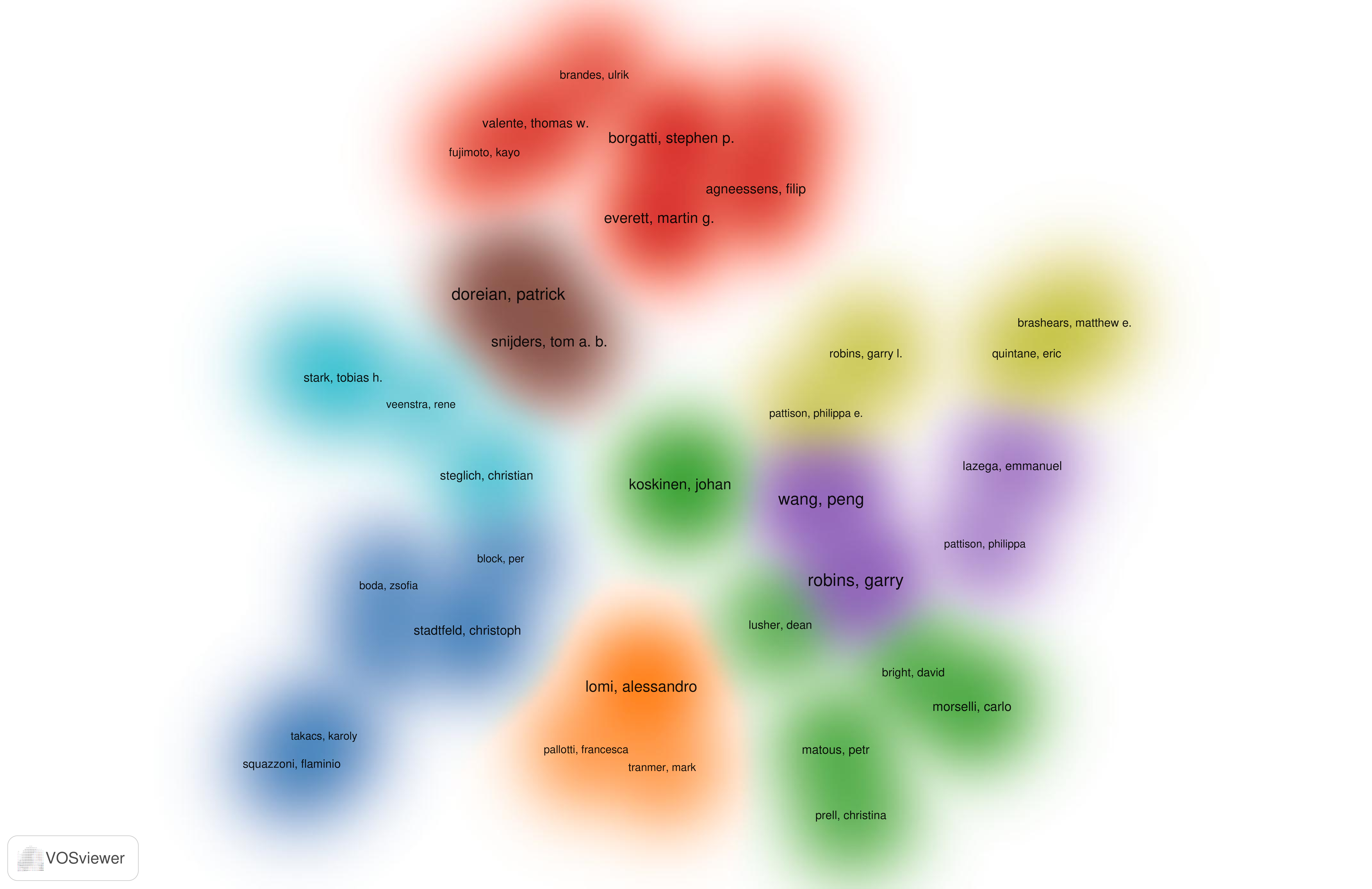}
			\caption{Co-authorship network density visualization}
			\label{fig.fig14}
		\end{figure}

		To assess the density and connectivity within the co-authorship network, we further analyzed the structural properties of the network. Figure \ref{fig.fig14} presents the co-authorship density map, where highly collaborative authors appear in denser regions. The intensity of colors in the visualization represents the concentration of co-authored works, with brighter regions indicating stronger collaboration patterns.
		
		The co-authorship network reveals several key insights into the collaborative structure of research in \textit{Social Networks}:
		
		\begin{itemize}
			\item \textbf{Prominent Authors:} The network highlights several influential authors, including \textit{Tom A.B. Snijders, Stephen P. Borgatti, and Garry Robins}, who exhibit strong co-authorship connections.
			\item \textbf{Clustered Communities:} The network is characterized by well-defined clusters, each representing a group of authors who frequently collaborate. These clusters correspond to distinct research subfields within network science.
			\item \textbf{Interdisciplinary Connections:} Some nodes serve as bridges between different research clusters, facilitating knowledge exchange across disciplines.
			\item \textbf{High-Density Regions:} The density visualization identifies research hubs where collaboration is particularly intense, suggesting influential research groups driving the field forward.
		\end{itemize}
		
		\subsubsection*{Network Science}
		
		To analyze the co-authorship patterns in the journal \textit{Social Networks}, we constructed a co-authorship network using bibliometric data. The visualization was generated using VOSviewer 1.6.20~\cite{vanEck2010vosviewer}, which clusters authors based on their collaborative relationships. Figure \ref{fig.fig13} illustrates the co-authorship network, where nodes represent individual authors, and edges denote co-authorship links. The size of each node corresponds to the author's publication volume, while edge thickness reflects the frequency of co-authorship~\cite{glanzel2004handbook, newman2004coauthorship}.

		\begin{figure}[htbp]
			\centering
			\includegraphics[width=\columnwidth]{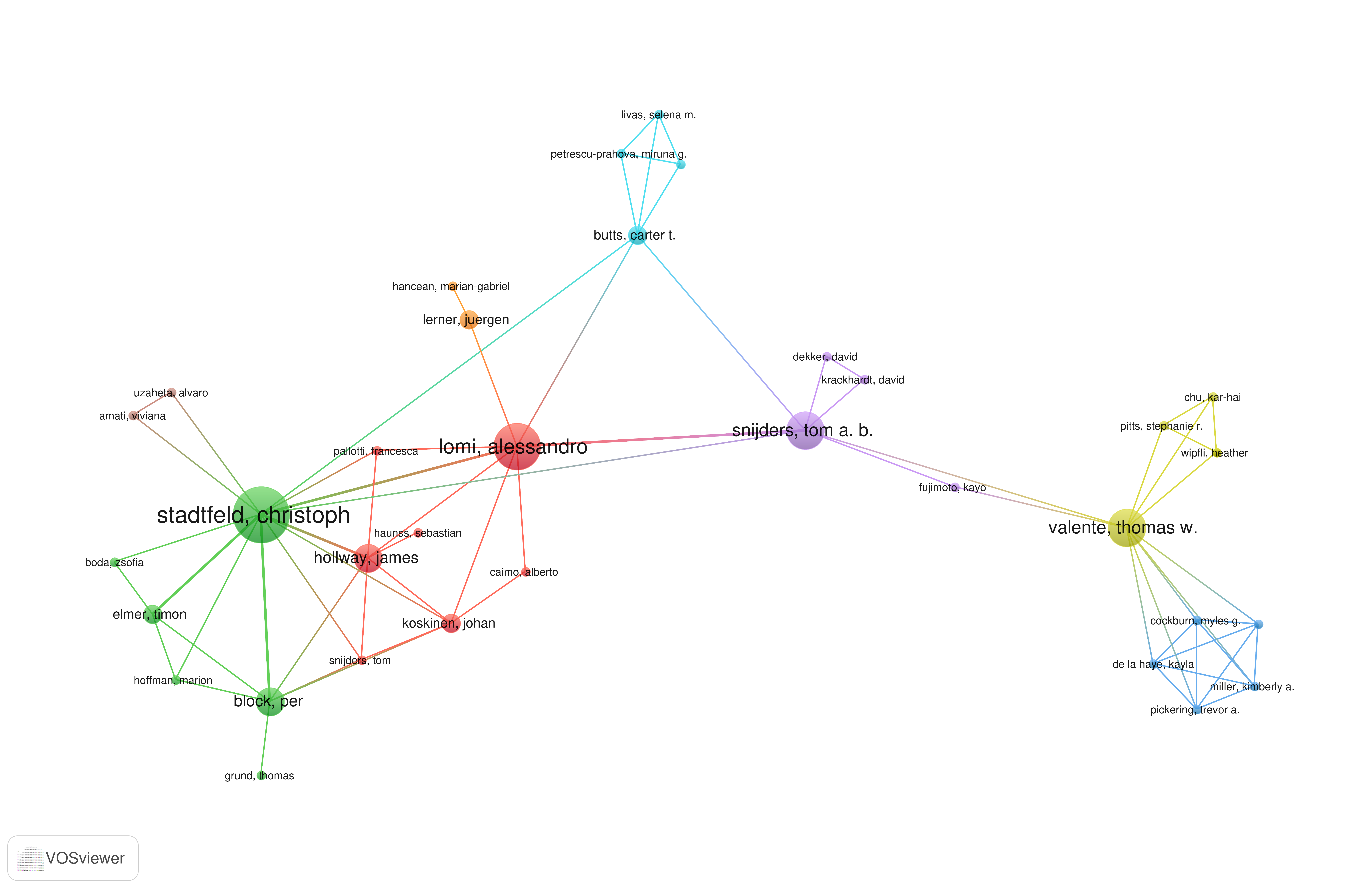}
			\caption{Co-authorship network visualization for \textit{Network Science}}
			\label{fig.fig15}
		\end{figure}
		
\begin{figure}[htbp]
			\centering
			\includegraphics[width=\columnwidth]{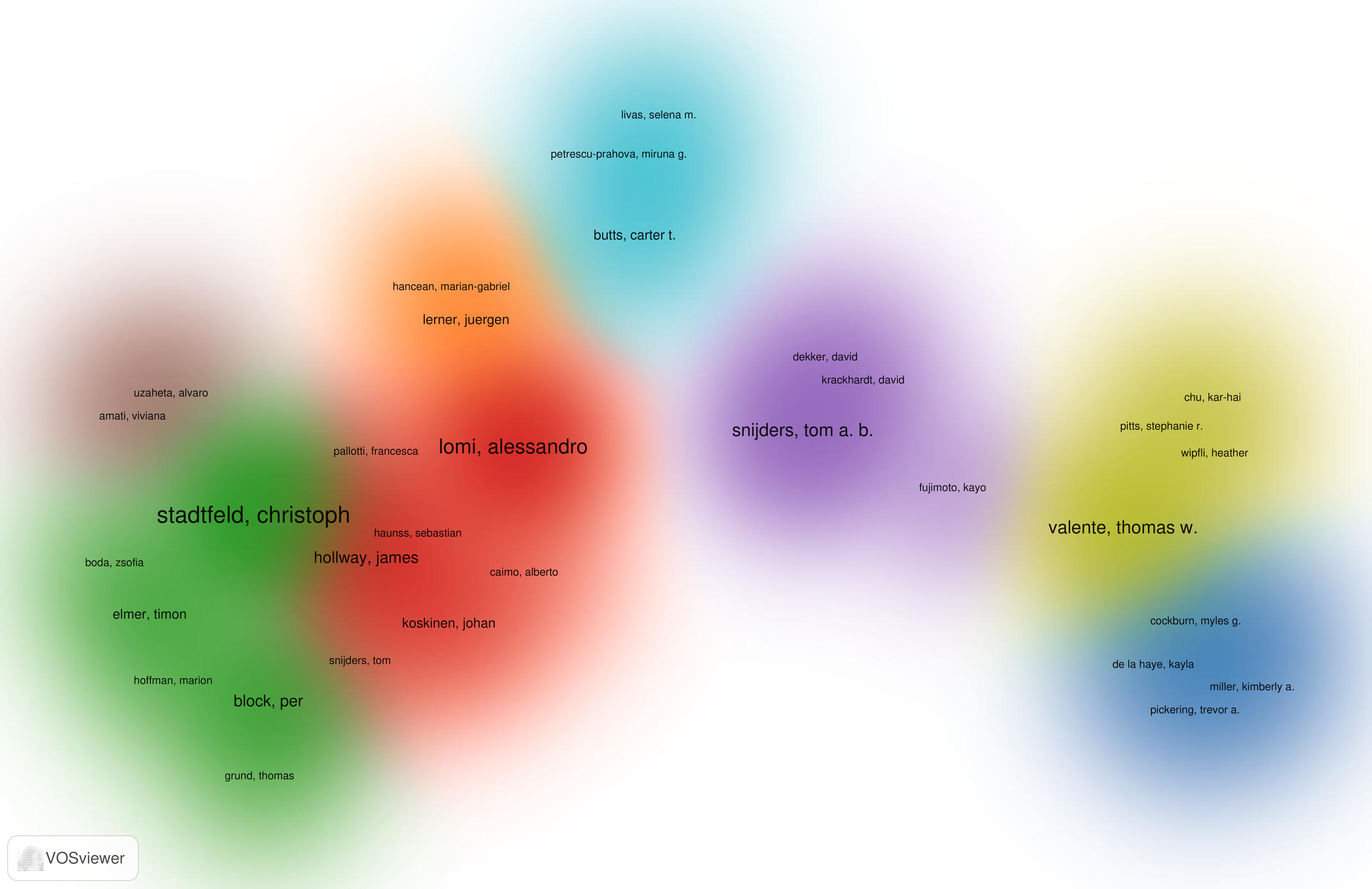}
			\caption{Co-authorship network density visualization for \textit{Network Science}}
			\label{fig.fig16}
		\end{figure}

		To further assess the network’s structural properties, we examined its density and connectivity. Figure \ref{fig.fig16} illustrates the co-authorship density map, highlighting regions of intense collaboration. The color gradient in the visualization signifies the concentration of co-authored works, with brighter areas indicating more frequent collaboration.
		
		Analysis of the co-authorship network yields several insights into collaborative research patterns within \textit{Network Science}:
		
		\begin{itemize}
			\item \textbf{Key Contributors:} The network underscores the prominence of several influential scholars, including \textit{Tom A.B. Snijders, Carter T. Butts, and Alessandro Lomi}, who maintain extensive co-authorship ties.
			\item \textbf{Distinct Research Groups:} The network is composed of well-defined clusters, each representing a set of researchers frequently collaborating within a particular subfield of network science.
			\item \textbf{Interdisciplinary Linkages:} Certain nodes function as bridges between clusters, facilitating the exchange of ideas across different research domains.
			\item \textbf{Areas of High Collaboration:} The density visualization pinpoints key research hubs where collaboration is particularly active, highlighting influential groups shaping the field.
		\end{itemize}
		
		\subsubsection*{Journal of Complex Networks}
		Also, for Journal of Complex Networks: 
		
		\begin{figure}[htbp]
			\centering
			\includegraphics[width=\columnwidth]{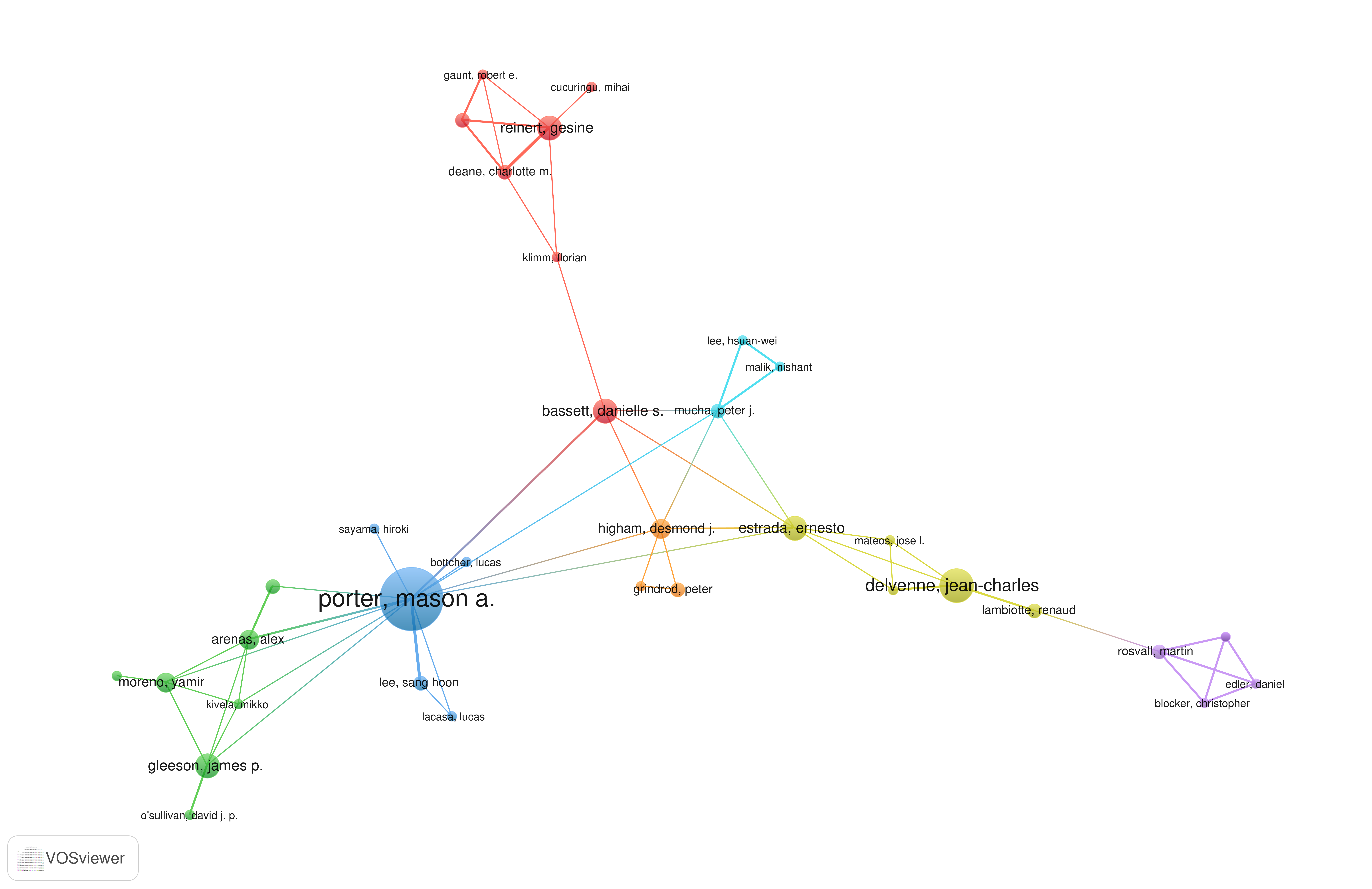}
			\caption{Co-authorship network visualization in the \textit{Journal of Complex Networks}.}
			\label{fig.fig12}
		\end{figure}
		
		To further analyze the structural characteristics of the co-authorship network, we examined the density of connections among researchers. Network density measures the overall level of collaboration in the academic community and reflects the extent to which authors frequently co-publish with one another~\cite{newman2001scientific, moody2004sociology, Barabasi2016}. High-density networks indicate strong interdisciplinary cooperation and frequent scholarly interactions, whereas lower-density networks suggest a more fragmented research landscape. This analysis provides deeper insights into the cohesion and collaborative dynamics within the field.

		\begin{figure}[htbp]
			\centering
			\includegraphics[width=\columnwidth]{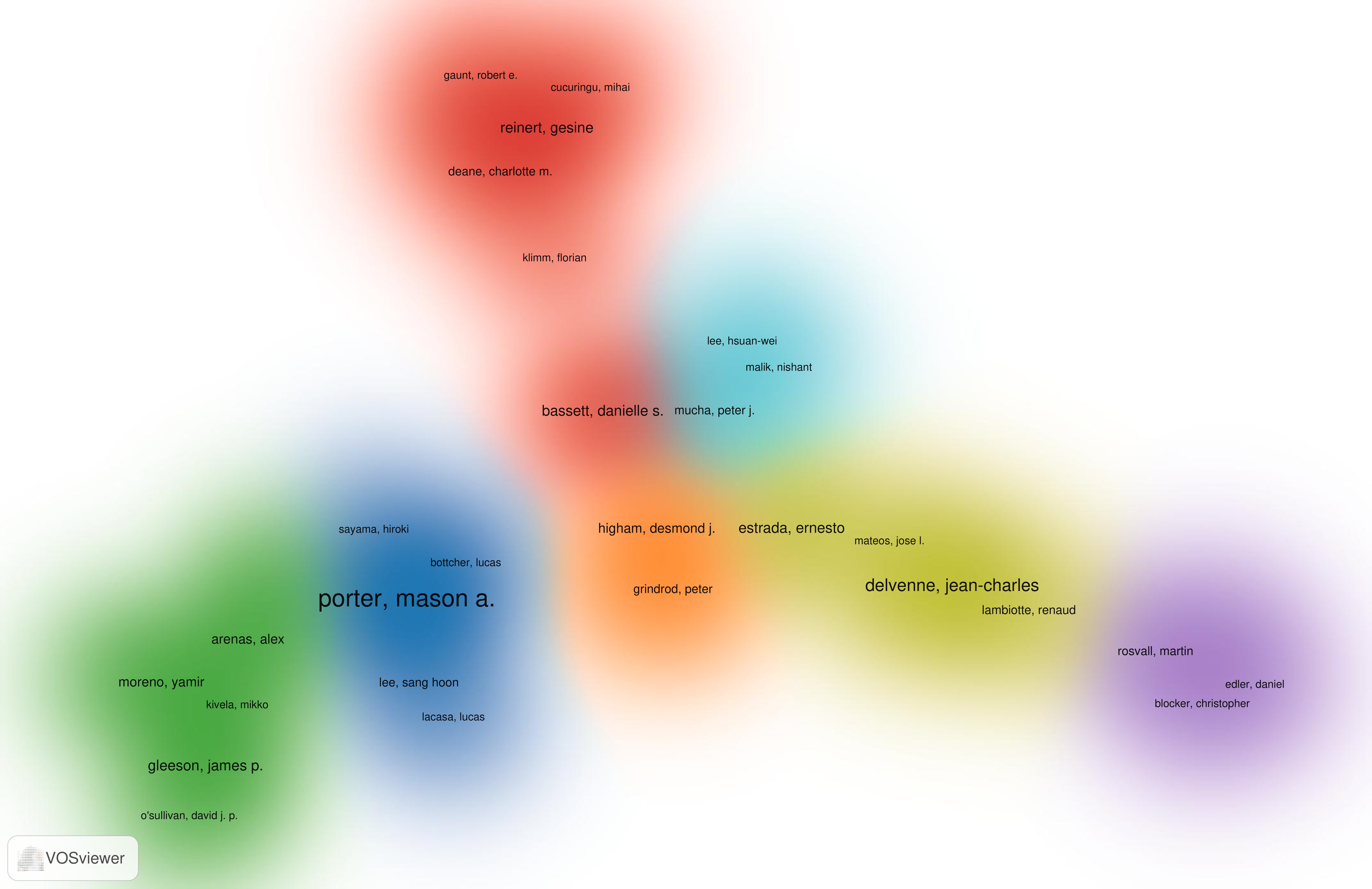}
			\caption{Co-authorship network density visualization in the \textit{Journal of Complex Networks}.}
			\label{fig.fig17}
		\end{figure}
		
		The analysis of the co-authorship network yields several key insights into the collaborative structure of research within the \textit{Journal of Complex Networks}:
		
		\begin{itemize}
			\item \textbf{Prominent Authors:} The network highlights several influential contributors, including Hiroki Sayama, David J.P. O’Sullivan, José L. Mateos, Nishant Malik, Hsuan-Wei Lee, and Lucas Lacasa, who exhibit extensive co-authorship connections.
			\item \textbf{Clustered Research Communities:} The co-authorship network is characterized by distinct clusters, each representing a group of researchers who frequently collaborate. These clusters likely correspond to specialized subfields within complex network science.
			\item \textbf{Interdisciplinary Collaborations:} Several authors serve as intermediaries connecting different research clusters, facilitating the exchange of knowledge across distinct domains.
			\item \textbf{High-Density Collaborative Hubs:} The density visualization highlights key research hubs where collaboration is particularly strong, indicating the presence of influential research groups shaping the direction of the field.
		\end{itemize}
		
		\section{Network Metrics}
		
		Network analysis is a key bibliometric technique for examining the structure of scholarly interactions by mapping relationships among authors, institutions, and research themes~\cite{newman2001scientific, Barabasi2002, Borgatti2009}. By constructing co-authorship, institutional collaboration, and citation networks, this approach offers valuable insights into the structural properties of academic discourse and the diffusion of knowledge~\cite{Otte2002, Liu2005}.
		
		To assess the impact of individual publications, we compute key network metrics, including degree centrality, betweenness centrality, and sigma scores. Table~\ref{table.tab5} highlights the most influential publications based on these metrics across the three journals analyzed.
		
		For accuracy, we employ NetworkX~\cite{SciPyProceedings_11} to calculate the network metrics. The results are detailed in Table~\ref{table.tab6}. We ranked the list based on a combined evaluation of all three centrality metrics.
		
		\begin{table*}[htbp]
			\centering
			\caption{Top 10 Publications in 3 Journals with Highest Centrality}
			\label{table.tab6}
			\resizebox{\textwidth}{!}{
				\begin{tabular}{lcccccc}
					\toprule
					Label & Degree & Degree Centrality & Betweenness Centrality & Closeness Centrality & Eigenvector Centrality & Journal \\
					\midrule
					Scott J, 2017, SOCIAL NETWORK ANALYSIS, V0, P0 & 12 & 0.007335 & 0.239656 & 0.156274 & 0.000088 & Social Networks \\
					BORGATTI SP, 1992, SOC NETWORKS, V14, P91, DOI 10.1016/0378-8733(92)90015-Y & 8 & 0.004890 & 0.178918 & 0.142706 & 0.000465 & Social Networks \\
					Block P, 2015, SOC NETWORKS, V40, P163, DOI 10.1016/j.socnet.2014.10.005 & 24 & 0.014670 & 0.108329 & 0.141205 & 0.000058 & Social Networks \\
					Wang P, 2013, SOC NETWORKS, V35, P96, DOI 10.1016/j.socnet.2013.01.004 & 24 & 0.014670 & 0.094689 & 0.154768 & 0.000206 & Social Networks \\
					Wasserman S, 1994, SOCIAL NETWORK ANAL, V0, P0 & 26 & 0.015892 & 0.079835 & 0.143585 & 0.000021 & Social Networks \\
					BURT RS, 1987, AM J SOCIOL, V92, P1287, DOI 10.1086/228667 & 24 & 0.014670 & 0.077956 & 0.109768 & 0.013629 & Social Networks \\
					Zappa P, 2015, ORGAN RES METHODS, V18, P542, DOI 10.1177/1094428115579225 & 18 & 0.011002 & 0.073588 & 0.153603 & 0.000134 & Social Networks \\
					DOREIAN P, 1987, SOC NETWORKS, V9, P89, DOI 10.1016/0378-8733(87)90008-6 & 17 & 0.010391 & 0.072486 & 0.120457 & 0.002516 & Social Networks \\
					BURT RS, 1983, APPLIED NETWORK ANAL, V0, P0 & 25 & 0.015281 & 0.071772 & 0.101817 & 0.039599 & Social Networks \\
					Burt RS, 1992, STRUCTURAL HOLES, V0, P0 & 17 & 0.010391 & 0.068141 & 0.134295 & 0.000098 & Social Networks \\
					Snijders TAB, 2017, ANNU REV STAT APPL, V4, P343, DOI 10.1146/annurev-statistics-060116-054035 & 8 & 0.020833 & 0.066634 & 0.088163 & 0.000004 & Network Science \\
					Block P, 2015, SOC NETWORKS, V40, P163, DOI 10.1016/j.socnet.2014.10.005 & 10 & 0.026042 & 0.066090 & 0.081605 & 0.000002 & Network Science \\
					Amati V, 2019, SOC NETWORKS, V57, P18, DOI 10.1016/j.socnet.2018.10.001 & 20 & 0.052083 & 0.058764 & 0.091121 & 0.000015 & Network Science \\
					Lomi A, 2012, SOC NETWORKS, V34, P101, DOI 10.1016/j.socnet.2010.10.005 & 9 & 0.023438 & 0.054830 & 0.074176 & 0.000007 & Network Science \\
					Leenders RTAJ, 2016, ORGAN PSYCHOL REV, V6, P92, DOI 10.1177/2041386615578312 & 5 & 0.013021 & 0.045654 & 0.084257 & 0.000003 & Network Science \\
					Brashears ME, 2018, SOC NETWORKS, V55, P104, DOI 10.1016/j.socnet.2018.05.010 & 6 & 0.015625 & 0.045566 & 0.078896 & 0.000000 & Network Science \\
					Bakshy E, 2015, SCIENCE, V348, P1130, DOI 10.1126/science.aaa1160 & 6 & 0.015625 & 0.036677 & 0.060877 & 0.000000 & Network Science \\
					Valente TW, 2012, SCIENCE, V337, P49, DOI 10.1126/science.1217330 & 13 & 0.033854 & 0.029537 & 0.065944 & 0.000007 & Network Science \\
					Salehi M, 2015, IEEE T NETW SCI ENG, V2, P65, DOI 10.1109/TNSE.2015.2425961 & 9 & 0.023438 & 0.026723 & 0.054533 & 0.000001 & Network Science \\
					Elmer T, 2019, BEHAV RES METHODS, V51, P2120, DOI 10.3758/s13428-018-1180-y & 10 & 0.026042 & 0.026261 & 0.087756 & 0.000004 & Network Science \\
					Allen-Perkins A, 2019, J STAT MECH-THEORY E, V2019, P0, DOI 10.1088/1742-5468/ab5700 & 3 & 0.005848 & 0.000000 & 0.131349 & 0.000047 & Journal of Complex Networks \\
					Miklós I, 2013, ELECTRON J COMB, V20, P0 & 2 & 0.003899 & 0.002155 & 0.077071 & 0.000000 & Journal of Complex Networks \\
					BORGATTI SP, 2011, ORGAN SCI, V22, P1168, DOI 10.1287/orsc.1100.0641 & 2 & 0.003899 & 0.000000 & 0.104677 & 0.000025 & Journal of Complex Networks \\
					Virtanen P, 2020, NAT METHODS, V17, P261, DOI 10.1038/s41592-019-0686-2 & 1 & 0.001949 & 0.000000 & 0.001949 & 0.000000 & Journal of Complex Networks \\
					Andjelkovic M, 2020, SCI REP-UK, V10, P0, DOI 10.1038/s41598-020-74392-3 & 7 & 0.013645 & 0.000013 & 0.080586 & 0.000000 & Journal of Complex Networks \\
					Rattana P, 2013, B MATH BIOL, V75, P466, DOI 10.1007/s11538-013-9816-7 & 1 & 0.001949 & 0.000000 & 0.002599 & 0.000000 & Journal of Complex Networks \\
					Nyberg A, 2015, J COMPLEX NETW, V3, P543, DOI 10.1093/comnet/cnv004 & 2 & 0.003899 & 0.000000 & 0.003899 & 0.000000 & Journal of Complex Networks \\
					van der Hofstad R, 2017, PHYS REV E, V95, P0, DOI 10.1103/PhysRevE.95.022307 & 2 & 0.003899 & 0.000000 & 0.003899 & 0.000000 & Journal of Complex Networks \\
					Asztalos A, 2012, EUR PHYS J B, V85, P0, DOI 10.1140/epjb/e2012-30122-3 & 1 & 0.001949 & 0.000000 & 0.119835 & 0.000093 & Journal of Complex Networks \\
					\bottomrule
				\end{tabular}
			}
		\end{table*}
		
		Table~\ref{table.tab6} presents the ten most central publications in \textit{Social Networks}, \textit{Network Science}, and \textit{Journal of Complex Networks}, ranked based on multiple centrality measures. These rankings offer a structural perspective on the most influential contributions within each journal, revealing key theoretical, methodological, and applied research trends. The centrality measures considered—degree centrality, betweenness centrality, closeness centrality, and eigenvector centrality—capture different aspects of a publication’s influence in the co-authorship and citation networks.
		
		The most central publications in \textit{Social Networks} are predominantly foundational works that have shaped the field of network analysis. Classical contributions such as those by Scott (2017), Borgatti (1992), and Wasserman (1994) \cite{Scott2017,Borgatti1992,wasserman1994social} remain highly central due to their widespread adoption. These studies introduced key concepts and methodological advancements that continue to influence contemporary research. The presence of Burt's work \cite{Burt1983,Burt1992,Burt1992} further underscores the lasting significance of structural hole theory in understanding social network dynamics. Additionally, more recent works such as Block \cite{Block2015} and Wang \cite{Wang2013} highlight the increasing role of statistical modeling in network science, reflecting the methodological evolution of the field. Other highly central publications, such as Zappa (2015) \cite{Zappa2015} and Doreian (1987) \cite{Doreian1987}, indicate the application of network models to organizational and structural research.
		
		In \textit{Network Science}, central publications primarily focus on computational and statistical methodologies. Works such as Snijders (2017) \cite{Snijders2017} and Block (2015) \cite{Block2015} exemplify the application of stochastic actor-based models, while Valente (2012) \cite{Valente2012} and Bakshy (2015) \cite{Bakshy2015} demonstrate the role of network diffusion and influence dynamics in large-scale systems. Additionally, Amati (2019) \cite{Amati2019} and Lomi (2012) \cite{Lomi2012} highlight the field’s emphasis on empirical network analysis using advanced statistical techniques. The presence of Salehi (2015) \cite{Salehi2015} and Elmer (2019) \cite{Elmer2019} suggests that \textit{Network Science} serves as a critical venue for interdisciplinary methodological developments. Compared to \textit{Social Networks}, which focuses more on empirical and theoretical work in sociology, \textit{Network Science} leans heavily toward computational approaches and large-scale network analysis.
		
		The highly central publications in \textit{Journal of Complex Networks} illustrate the journal’s interdisciplinary focus. While the overall centrality values of these works are lower than those in \textit{Social Networks} and \textit{Network Science}, they highlight the diverse applications of network science across multiple domains. Studies such as Allen-Perkins (2019) \cite{AllenPerkins2019} and Borgatti (2011) \cite{Borgatti1992} demonstrate how network methodologies are adapted to different research areas, while works like Andjelkovic (2020) \cite{Andjelkovic2020} and Nyberg (2015) \cite{Nyberg2015} emphasize the integration of network models in biological and physical systems. The presence of Asztalos (2012) \cite{Asztalos2012} and van der Hofstad (2017) \cite{vanDerHofstad2017} suggests a growing interest in applying statistical physics to complex networks, reinforcing the journal’s role in bridging traditional network science with broader scientific disciplines.
		
		A comparative analysis of centrality measures provides us with additional insights into the structural positioning of these publications. Works appearing in \textit{Social Networks} tend to exhibit higher degree centrality, indicating frequent citation and recognition within the academic community. Publications with high betweenness centrality, such as Scott (2017) \cite{Scott2017} and Borgatti (1992) \cite{Borgatti1992}, serve as key bridging studies that connect distinct research areas. In contrast, studies with high closeness centrality, including Zappa (2015) \cite{Zappa2015} and Doreian (1987) \cite{Doreian1987}, are structurally positioned to influence a broad range of works due to their network proximity to other publications. Eigenvector centrality, which reflects influence within a citation cluster, highlights the prominence of works such as Burt (1983, 1992) \cite{Burt1983,Burt1992} and Amati (2019) \cite{Amati2019}, suggesting their importance within their respective research communities.
		
		The centrality rankings provide a structured view of the most influential works across three major journals in network science. While \textit{Social Networks} primarily features foundational theoretical and methodological contributions, \textit{Network Science} emphasizes computational and statistical modeling approaches. \textit{Journal of Complex Networks}, in contrast, displays a broader disciplinary focus, incorporating network-based methodologies across various scientific domains. These findings highlight the evolving intellectual landscape of network science and offer insights into how different research traditions contribute to the field’s development. 
		
		\section{Conlusion}
		
		This bibliometric study of social network research across multiple journals reveals the intricate and evolving structure of this academic domain. By examining co-authorship networks, citation patterns, and key centrality measures, we uncover not only the dominant figures and influential publications but also the underlying intellectual currents shaping the field. The presence of well-connected authors and tightly clustered research communities indicates that social network research has matured into a structured discipline with distinct methodological and theoretical foundations.
		
		One of the most striking observations is the persistence of foundational works maintaining their relevance over time, demonstrating that certain conceptual and methodological contributions continue to serve as the backbone of network analysis. At the same time, the emergence of new high-impact publications suggests that the field is undergoing expansion, incorporating interdisciplinary perspectives and computational advancements. The increasing prominence of multilayer and dynamic network models reflects a shift toward more complex representations of social structures, driven by the need to account for temporal and contextual variations in relational data.
		
		The patterns of collaboration further highlight the dynamics of knowledge production in this field. Highly centralized networks, where a few scholars act as bridges between research clusters, suggest that intellectual leadership is concentrated among a small but influential group of researchers. These bridging scholars facilitate the diffusion of ideas, integrating diverse approaches and fostering methodological innovation. Conversely, the presence of insular clusters hints at specialized subfields that develop independently, potentially limiting cross-fertilization between research traditions.
		
		The study also reveals an increasing reliance on computational methodologies, particularly in areas where traditional social science approaches intersect with machine learning, big data analytics, and network inference techniques. This methodological convergence signifies a transformation in how network science is conceptualized and applied, extending beyond classical sociological and statistical approaches to incorporate more predictive and algorithmic paradigms.
		
		Looking ahead, the trajectory of social network research will likely be shaped by the growing influence of digital platforms, the proliferation of large-scale networked data, and the increasing demand for network-based interventions in fields such as public health, policy-making, and organizational behavior. The challenge lies in balancing theoretical rigor with methodological innovation, ensuring that as the field expands, it retains coherence and conceptual depth. By continuously refining its analytical frameworks and embracing interdisciplinary perspectives, social network research is poised to remain a critical lens for understanding the complexities of human connectivity in an increasingly networked world.

	\bibliographystyle{apalike}
	\bibliography{ref}
	
\end{document}